\documentclass[fleqn,10pt]{wlscirep}
\usepackage{bm}

\usepackage{comment}

\usepackage[utf8]{inputenc}
\usepackage[T1]{fontenc}
\usepackage{siunitx}
\usepackage[normalem]{ulem}

\usepackage[graphicx]{realboxes}

\DeclareSIUnit{\myeuro}{\text{€}}
\title{Robust design of bicycle infrastructure networks}

\author[1,*]{Christoph Steinacker}
\author[2]{Mads Paulsen}
\author[1,+]{Malte Schröder}
\author[2,+]{Jeppe Rich}

\affil[1]{Chair of Network Dynamics, Center for Advancing Electronics Dresden (cfaed) and Institute of Theoretical Physics, TUD Dresden University of Technology, 01062 Dresden, Germany}
\affil[2]{Transportation Science Division, Department of Technology, Management and Economics, Technical University of Denmark, 2800 Kgs. Lyngby, Denmark}

\affil[*]{christoph.steinacker@tu-dresden.de}

\affil[+]{these authors contributed equally to this work}

\begin{abstract}
Promoting active mobility like cycling relies on the availability of well-connected, high-quality bicycle networks. However, expanding these networks over an extended planning horizon presents one of the most complex challenges in transport science. This complexity arises from the intricate interactions between infrastructure availability and usage, such as network spillover effects and mode choice substitutions. In this paper, we approach the problem from two perspectives: direct optimization methods, which generate near-optimal solutions using operations research techniques, and conceptual heuristics, which offer intuitive and scalable algorithms grounded in network science. Specifically, we compare direct welfare optimization with an inverse network percolation approach to planning cycle superhighway extensions in Copenhagen. Interestingly, while the more complex optimization models yield better overall welfare results, the improvements over simpler methods are small. More importantly, we demonstrate that the increased complexity of planning approaches generally makes them more vulnerable to input uncertainty, reflecting the bias-variance tradeoff. This issue is particularly relevant in the context of long-term planning, where conditions change during the implementation of the planned infrastructure expansions. Therefore, while planning bicycle infrastructure is important and renders exceptionally high benefit-cost ratios, considerations of robustness and ease of implementation may justify the use of more straightforward network-based methods.
\end{abstract}

\begin{document}

\flushbottom
\maketitle

\thispagestyle{empty}

\section{Introduction}

Promoting active mobility, like walking and cycling, is important for achieving sustainable mobility. However, this is only possible with a sufficiently developed infrastructure. Specifically for cycling, demand strongly relies on the quality and quantity of the available bicycle infrastructure \cite{Creutzig2016_UrbanInfrastructureClimate, Buehler2011_CyclingToWork, Fosgerau2023_InducedDemand}. Due to its low costs and space requirements compared to other modes of transport \cite{Buczynski2021_costcycling, NelloDeakin2019}, there is growing evidence of large, positive welfare benefits of bicycle infrastructure for society \cite{Rich2021, Paulsen2023_SocietallyOptimalExpansion, Paulsen2024_WelfareOptimalExpansionInducedDemand, Saelensminde2004, Chapman2018, Whitehurst2021}. 

\begin{figure}[hbt!]
    \centering
    \includegraphics[scale=0.95]{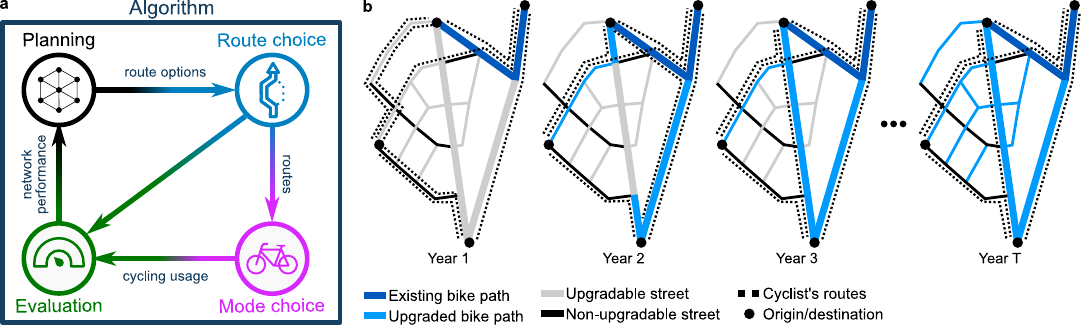}
    \caption{The network expansion problem. (a) Designing efficient infrastructure network expansions is a complex problem due to interactions and feedback effects between the available and planned infrastructure, the possible routing options, the resulting cycling demand, and the evaluation of the network quality. (b) Sketch of bike path network extensions over time (left to right). Cyclists prefer to travel along bike paths (blue lines) but avoid large streets without bike paths (thick gray lines). As new bike paths are added (light blue), the cyclists change their routes (black dotted lines) to use the new routing options. This in turn affects the importance of all other existing and potential bike paths in the network, and thus changes which bike paths should be constructed next.} 
    \label{fig:FIG1_NetworkExpansion}
\end{figure}

However, designing efficient network expansions over time represents one of the most challenging problems in urban planning \cite{Leblance1975, Li2017}. The complexity of these network design or network expansion problems \cite{Yang1998_ModelsDevelopments, Magnanti1984} stems from their dynamic nature, where investments in one period affect the usage of the network and the benefits of additional investments in the future (Fig.~\ref{fig:FIG1_NetworkExpansion}). These feedback effects include changes in the routes of cyclists and spillovers from other transport modes as cycling becomes faster and more convenient. The problem presents itself as a dynamic optimization problem with a structure similar to the one presented in the seminal work of Bellman \cite{Bellman1952}. Compared to complex network design problems for multimodal or congested transport, bicycle infrastructure planning does not typically include direct interactions between individual travelers. Nonetheless, the problem remains highly complex \cite{Johnson1978_ComplexityNetworkDesign} due to the strong interactions and indirect feedback loops between available infrastructure, network quality, cycling demand, and route choice (Fig.~\ref{fig:FIG1_NetworkExpansion}) and cannot be solved without substantial simplifications of the underlying problem. This has resulted in a divide between two algorithmic approaches to addressing the problem. 

One school addresses the problem from the classical operations research perspective, typically framing it as a mathematical program to optimize various objectives. These objectives can range from optimizing travel time \cite{Luathep2011} or distance and safety indicators \cite{Lim2022} over minimizing cost and investments \cite{Duthie2014} to maximizing connectedness and directness of routes expressed in a set-covering mixed-integer linear programming (MILP) \cite{Ospina2022}. Many approaches combine multiple quality measures in a weighted multiple criteria function to represent the costs and benefits of all stakeholders involved \cite{Hosseininasab2018, Owais2016_MultiObjective} or employ Pareto-optimization to identify different feasible networks, for example based on the trade-off in travel times between bike and car traffic \cite{Ballo2023_EbikeCity}. In recent years, a new class of operations research models has emerged, which more closely follows traditional cost-benefit analyses and aims to optimize a general welfare function expressed in terms of the net present value (NPV) of the infrastructure \cite{Xinxin2022, Paulsen2023_SocietallyOptimalExpansion, Paulsen2024_WelfareOptimalExpansionInducedDemand}. The net present value models also combine multiple criteria, all weighted into a common monetary base, to evaluate the total societal benefit of network extensions. In the reference model applied in this paper, the multi criteria welfare function includes consumer surplus effects for cyclists, reduced healthcare costs for society due to the benefits of cycling, and the cost of investments and infrastructure maintenance \cite{Paulsen2024_WelfareOptimalExpansionInducedDemand}. Typically, to solve these problems on a large scale, various simplifications are required to enable the evaluation of the objective function or the optimization process itself, such as heuristic search of the possible extension pathways, aggregation and linearization techniques to simplify complex computations, or treating the optimization as a Markov chain problem and iteratively finding the next expansion steps based on the current network state instead of computing the all expansion steps at once. Overall, the operations research perspective focuses more strongly on devising efficient solution methods and achieving network extension plans as close to optimal as possible.

The second school has grown out of the network science community and often approaches the problem more conceptually. These network science approaches analyze large-scale network data \cite{Viero2024_BikeDNA} and employ simplified heuristic models of network expansion based on conceptual quality measures including connecting fragmented components \cite{NateraOrozco2020_DatadrivenStrategiesOptimal, Vybornova2022_MissingLinkDetection, Hwang2024}, directness \cite{Olmos2020_DataScienceFramework, Mahfouz2023_RoadSegmentPrioritization}, effective shortest path distances \cite{Steinacker2022_DemandDrivenDesign} and coverage of the network \cite{Szell2022_GrowingUrbanBicycle,  Akhand2021_BicycleNetworkDesign}.
A recent model taken as a reference in this paper, for example, generates a sequence of infrastructure networks by iteratively deconstructing a fully expanded network based on a flexible network evaluation scheme \cite{Steinacker2022_DemandDrivenDesign}. Overall, unlike the operations research perspective, network science-inspired approaches prioritize understanding common structural characteristics of efficient networks over identifying the exact optimal solution. This understanding can, for instance, inform heuristic search methods or help assess the quality of existing networks with interpretable measures. 

How do these two communities' network planning strategies and suggested network expansions compare when applied to the same network expansion problem? 
Here, we aim to bridge this research gap by directly comparing the suggested expansion steps from a state-of-the-art mathematical programming model recently published by Paulsen \& Rich \cite{Paulsen2023_SocietallyOptimalExpansion, Paulsen2024_WelfareOptimalExpansionInducedDemand} to a dynamic percolation model suggested by Steinacker \emph{et~al.} \cite{Steinacker2022_DemandDrivenDesign} for planning the cycling superhighway network of Copenhagen with the same underlying network and routing data. Although the analysis of the results of the mathematical programming model indicates a small optimality gap, its practical implementation remains challenging due to its complexity and reliance on various inputs that are difficult to predict accurately. This raises concerns about its robustness when inputs are uncertain. By contrast, the simpler strategy proposed by \cite{Steinacker2022_DemandDrivenDesign} will have a larger optimality gap. However, it is easier to implement and benefits from a modular structure, enhancing its adaptability to diverse data availability settings. Based on the comparison of these two approaches, this paper presents three main contributions: 
Firstly, we show that the dynamic percolation model \cite{Steinacker2022_DemandDrivenDesign} is extendable to a wide range of objective functions, including an approximation of the social welfare function used in the mathematical programming model \cite{Paulsen2023_SocietallyOptimalExpansion}. 
Secondly, if the underlying data is known, the different expansion strategies render broadly similar overall performance with quality differences of less than $\SI{9}{\percent}$. Differences predominantly show as small variations in the specific ordering and the geographical distribution of investments. 
Thirdly, the more complicated models, here exemplified by the direct optimization model with a complex welfare function, tend to be more susceptible to the uncertainty of input data. This exemplifies the bias-variance trade-off in predictive planning and suggests that more complex models do not necessarily render more reliable predictions in practice.

\newpage

\section{Network setting and planning approaches}

\subsection{Copenhagen cycle superhighway network}
\begin{figure}[h]
    \centering
    \includegraphics[scale=0.95]{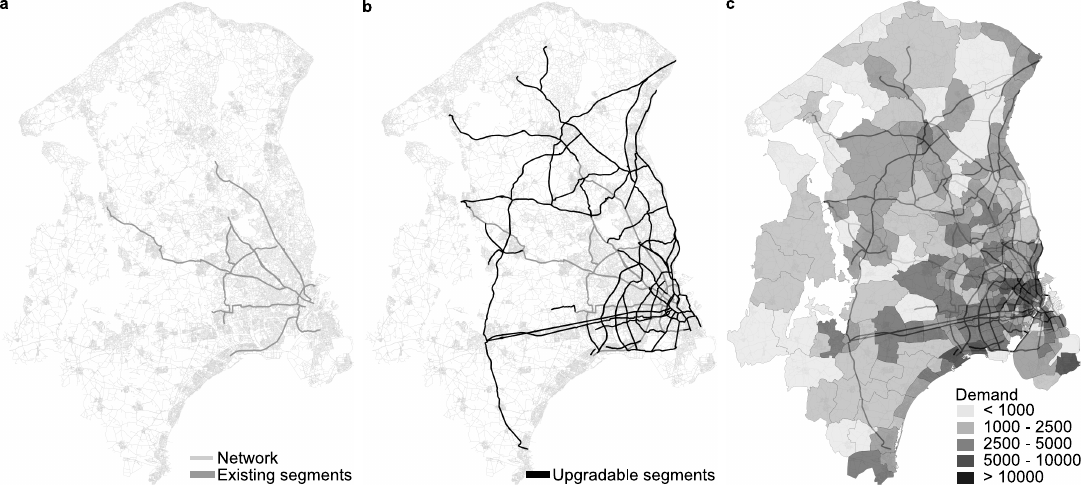}
    \caption{
    Cycle superhighway network in the greater Copenhagen area.
    (a) Existing infrastructure. Cyclists travel on all links of the network, representing streets and bike paths (light gray) and existing cycle superhighways (dark gray). 
    (b) Potential extensions to the cycle superhighway network. Each of the $202$ segments of the planned cycle superhighway (black) includes multiple edges of the underlying network.
    (c) Cycling demand. Demand is distributed over $258$ origin and destination zones, with the trips starting and ending at the network node closest to the centroid of each zone for a total of $52\,808$ origin-destination-pairs with non-zero demand. The shading denotes the number of cycling trips starting in each zone in the current network state (see Methods for details).
    }
    \label{fig:FIG2_CopenhagenSetting}
\end{figure}

Planning of the Copenhagen cycle superhighway network constitutes a large-scale network planning application. The existing network consists of $\SI{460}{\km}$ of cycle superhighways, mainly in the urban area of Copenhagen (Fig.~\ref{fig:FIG2_CopenhagenSetting}a). Planned extensions of the network include $S = 202$ additional segments with a total length of $\SI{1876}{\km}$ across the greater area of Copenhagen (Fig.~\ref{fig:FIG2_CopenhagenSetting}b). A before-and-after study by Skov-Petersen \emph{et~al.} \cite{SkovPetersen2017_EffectsSuperCycleHighways} of Copenhagen's supercycle highway infrastructure interventions has previously demonstrated the significant benefits of network upgrades and their positive impact on cycling demand, thus highlighting the importance of planning future network extensions.

In our model, each planned segment upgrade encompasses multiple edges that either need to be upgraded from the underlying street network or represent entirely new connections to be added to the network. Cyclists may travel on all links in the street and bike path network, preferentially using faster, more comfortable routes along cycle superhighways and bike paths. Here, we employ a simplified shortest path route choice model with empirical velocities to efficiently enable repeated routing calculations as the network changes during the planning simulations. The demand for cycling is modelled as a zone-based mode choice model with $52\,808$ origin-destination pairs (Fig.~\ref{fig:FIG2_CopenhagenSetting}c) further divided into nine cyclist types separated by fitness and the type of bicycle \cite{Hallberg2021_CycleSuperHighway}. See Methods for more details on the model setting and parameters.

We consider the expansion of the cycle superhighway network over a planning horizon of $50$ years. Individual segments are added in order of priority, with an annual budget of $B = 7.5$ million EUR for construction and maintenance of the network, carrying over the remaining funds to the next year. In all scenarios, the evaluation period is long enough to allow all segments to be constructed with the total allocated budget.

\newpage

\subsection{Network planning} 
In a world of perfect information, having more input data can improve models by providing a more detailed representation of objectives and outcomes. However, these added details come at a cost -- they increase computation time and demand more effort in data collection and projection. What is often overlooked is that such optimization models are subject to the bias-variance trade-off \cite{Kohavi1996}, a concept from statistics and machine learning. This trade-off suggests that while complex models may excel at replicating known data, they are often poor at making accurate predictions on new, unseen data. The comparison in this paper provides a compelling example of this trade-off, showing that while a highly detailed optimization approach performs well with perfect information, it struggles when input data is uncertain. Below we consider two extremes, one represented by the all-inclusive direct optimization approach and one by a simple bikeability index.  

\subsubsection{Network quality measures}
\paragraph{Net present value}
We evaluate the societal benefits of the cycle superhighway network in terms of its cumulative net present value (NPV) over the 50-year planning horizon. The net present value combines the direct benefits to the individual cyclists captured by the consumer surplus (CS), reduced societal costs due to health benefits (HB), the built segments' construction cost (CC) and maintenance cost (MC), and the remaining scrap value (SV) at the end of the 50-year planning horizon \cite{Paulsen2024_WelfareOptimalExpansionInducedDemand}. The individual terms are explained in Tab.~\ref{tab:npv_terms} and more details are provided in the Methods section. Together, the net present value 
\begin{align} \label{eq:NPV_main}
    \mathrm{NPV} &= \mathrm{CS} + \mathrm{HB} - \mathrm{CC} - \mathrm{MC} + \mathrm{SV}
\end{align}
represents the cumulative societal benefits and costs of the network extensions.

\begin{table}[h]
    \centering
    \renewcommand{\arraystretch}{1.3}
    \begin{tabular}{|l|p{10cm}|}
        \hline
        \textbf{Term} & \textbf{Explanation} \\
        \hline
        $\mathrm{CS}$ (Consumer Surplus) & Measures the effective travel time savings for cyclists due to the planned expansion compared to the baseline network. Represents the individual economic benefits of improved infrastructure. \\
        \hline
        $\mathrm{HB}$ (Health Benefits) & Measures the positive health effects of increased total distance cycled by the population due to induced demand compared to the baseline network. Represents reduced healthcare costs and overall societal well-being improvements. \\
        \hline
        $\mathrm{CC}$ (Construction Cost) & Measures the financial investment required to build the new cycle superhighway network segments. This is a one-time infrastructure cost. \\
        \hline
        $\mathrm{MC}$ (Maintenance Cost) & Measures the ongoing financial expenditures required to maintain the infrastructure over the 50-year planning horizon. \\
        \hline
        $\mathrm{SV}$ (Scrap Value) & Measures the discounted monetary value of the constructed infrastructure remaining at the end of the planning period. Represents the residual worth of assets after 50 years. \\
        \hline
    \end{tabular}
    \caption{Summary of NPV components and their explanations}
    \label{tab:npv_terms}
\end{table}

Typically, the travel time savings measured by the consumer surplus make up about one-third of the positive contribution to the net present value. Health benefits contribute to the remaining two-thirds. The negative contributions of construction and maintenance costs for bicycle infrastructure are comparatively small, amounting to less than $\SI{10}{\percent}$ of the absolute net present value. However, due to a limited budget, construction and maintenance costs strongly affect which and how many cycle superhighway segments are added to the network each year. For Eq.~\ref{eq:NPV_main} to be exact, the inputs need to be known precisely and deterministically for the entire investment horizon. If, on the other hand, inputs are considered uncertain, these errors will propagate over the years and render much less robust solutions which could perform worse compared to simpler variants. 

\newpage

\paragraph{Bikeability}
The bikeability index \cite{Steinacker2022_DemandDrivenDesign} is an example of a simpler network quality measure representing a non-monetary valuation based on improving cycling travel time and demand in the network. While it does not represent the monetary investment or full societal benefit of the network, it serves as a simple comparison of the efficiency of a bike path network in enabling direct travel and forms the basis of the network percolation approach introduced in Steinacker \emph{et~al.} \cite{Steinacker2022_DemandDrivenDesign}. To define the bikeability, we first define a loss function $L \ge 0$ for a single trip $\omega$ with travel time $\tau$ in the current network $G$ as the integral under the demand curve $n(\omega, \tau)$,
\begin{align} \label{eq:trip_lossfunction_main}
    L(\omega, G) &= \int_{0}^{\tau} n(\omega, \tau^\prime) \, \mathrm{d} \tau^\prime \,,
\end{align}
The loss function decreases when new bike paths decrease the travel time $\tau$ of the trip, with a stronger effect the larger the number $n(\omega, \tau)$ of cyclists on the trip. By summing up the loss functions $L(\omega,G)$ for all individual trips $\omega$, we obtain a network-wide loss function 
\begin{align} \label{eq:network_lossfunction_main}
    L(G) = \sum_{\omega} L(\omega, G) \,.
\end{align}
To evaluate the progress from the base network $G_\mathrm{base}$ to the fully upgraded network $G_\mathrm{full}$, we define the bikeability $B$ of an intermediate cycle superhighway network $G$ as the normalized improvement of the network-wide loss function $L$,
\begin{align} \label{eq:bikeability_main}
    B(G) &= \frac{L\left(G_\mathrm{base}\right) - L\left(G\right)}{L\left(G_\mathrm{base}\right) - L\left(G_\mathrm{full}\right)} \,,
\end{align}
such that $B(G_\mathrm{base}) = 0$ for the base network and $B(G_\mathrm{full}) = 1$ for the fully upgraded network \cite{Steinacker2022_DemandDrivenDesign} (see also Supplementary Note 1, including Fig.~\ref{fig:FIGS1_BikeabilityLossFunction}).

\newpage

\subsubsection{Network optimization strategies}

\paragraph{Direct optimization}
Direct optimization is, by nature, impractical for large-scale problems with extensive geographical coverage and long planning periods. As a result, applied solution methods are often either based on probabilistic sampling, e.g. simulated annealing \cite{Gastner2006_OptimalDesignSpatial}, genetic algorithms \cite{Hosseininasab2018}, or simplify the optimization problem into set-covering problems \cite{Ospina2022, Duthie2014}. As a reference approach in this paper, we employ iterative batched optimization of newly constructed cycle superhighway segments for each year of the planning period to maximize the net present value at the end of the planning period \cite{Paulsen2024_WelfareOptimalExpansionInducedDemand}. The method solves a mathematical program in each time period to determine the most societally profitable segments. If no such segments exist, no segments are constructed.

Additionally, we test a simplified greedy optimization approach based on a linear interpolation of cyclists' route choices between the original and the fully upgraded network. This greedy approach iteratively constructs the entire network by sequentially adding the most valuable cycle superhighway segment regarding its approximated net present value contribution per construction cost. The linearized evaluation of the objective function avoids the explicit calculation of the routes during the optimization, thus enabling a faster solution to the optimization problem. However, evaluating the quality of the resulting networks still requires the calculation of the routes of all cyclists in each network state. 

Expanding the network from the base state, both optimization algorithms explicitly include time-dependent information in the evaluation, such as budgets and discounting future benefits to extrapolate the value of the objective function at the end of the planning period. The methodology underlying both optimization approaches is described further in the Methods section below.

\paragraph{Dynamic backward percolation}
Percolation describes the emergence or breakdown of macroscopic connectivity in networks as links are added or removed. Dynamic percolation methods have initially been suggested to reveal the community structure of a network by iteratively removing links with the highest betweenness to disconnect the network as efficiently as possible \cite{Newman2004_FindingEvaluatingCommunity}. Similar approaches have since been applied to understand the structure of aviation networks by iteratively removing unprofitable connections \cite{Verma2014_WorldAirlineNetwork}. 

\begin{figure}[h]
    \centering
    \includegraphics[scale=0.95]{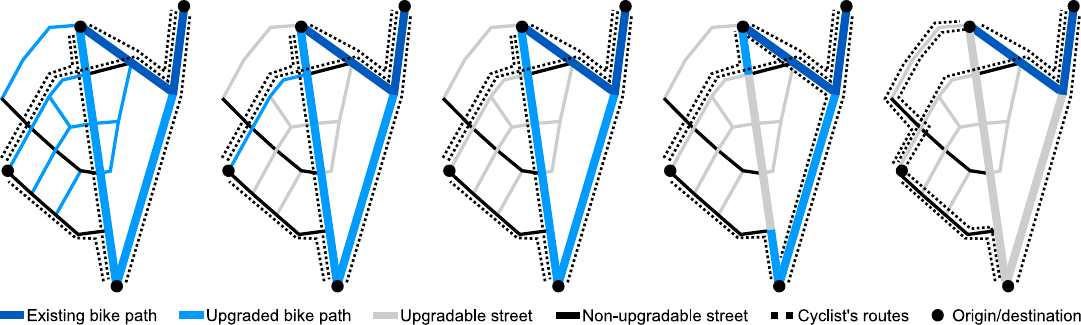}
    \caption{Dynamic backward percolation \cite{Steinacker2022_DemandDrivenDesign}.
    Thick lines represent main streets, thin lines represent residential streets. The color denotes the type of bike infrastructure: (gray and black) streets without a bike path, (dark and light blue) existing and newly constructed bike paths, respectively. Starting from a network where every upgradable street (gray) is equipped with a bike path (light blue), we first remove all unused bike paths given the demand distribution and the routes of all cyclists (dotted lines). We then iteratively remove the least important newly constructed bike path and recompute the routes of all affected cyclists. The process ends when all newly constructed bike paths have been removed and only the existing infrastructure (dark blue) remains. Reversing the order of bike path removal provides a prioritization for the bike path network extensions. 
    }
    \label{fig:FIG3_DynamicsInversePercolation}
\end{figure}

Recently, a bike path planning algorithm based on backward percolation was presented \cite{Steinacker2022_DemandDrivenDesign}. Starting from a fully upgraded network, the algorithm iteratively removes the least important bike path, generating a sequence of extensions in inverse order (Fig.~\ref{fig:FIG3_DynamicsInversePercolation}). This inverse approach avoids reinforcing inefficient route choices of cyclists in an incomplete network (e.g., taking detours to avoid busy streets without a bike path). In contrast, static or forward percolation approaches often lock cyclists in suboptimal routes when selecting network extensions based only on the route choices in the sparse base network of bike paths \cite{Olmos2020_DataScienceFramework, Szell2022_GrowingUrbanBicycle}. This is especially detrimental when planning network extensions with insufficient infrastructure.

The dynamic nature of the approach means that the routes of all cyclists are recomputed after a bike path has been removed. The algorithm thus automatically tracks the routes and demand as a function of the current bike path network, enabling online evaluation of the network quality. In addition, this makes the approach highly flexible and allows for the computation of almost arbitrary importance measures and route choice models. However, dense demand or complex models may result in long computation times. Additionally, due to the inverse construction, the order of extensions is only fixed once the algorithm has finished. Since this order is crucial to determine the cumulative maintenance cost and required budget, it is impossible to explicitly compute time-dependent evaluation functions like the net present value during the planning simulation. 

In this paper, we test this backward percolation approach with three different importance measures for the segments that represent successively more accurate approximations of their contribution to the net present value of the network. The first measure $Q_\mathrm{pen}(s)$ quantifies the weighted travel time penalty for cyclists when removing a segment $s$ relative to the length of the segment. The second measure $Q_\mathrm{stat}$ is derived from an explicit linearization of the change in consumer surplus relative to the construction costs of the segment, again assuming static routes and demand. Finally, assuming only static routes but allowing dynamic demand, the third measure $Q_\mathrm{dyn}$ is derived from a linear approximation of the consumer surplus and health benefits, capturing the majority of the change in net present value. All measures are explained in more detail in the Methods, derivations are provided in Supplementary Note 2 (including Fig.~\ref{fig:FIGS2_PenaltyModel}) and Supplementary Note 3, respectively.

\section{Results}
In the following analysis, we evaluate the performance of the different network planning approaches by focusing on four aspects: First, we assess the overall societal network performance. Next, we examine the geographical distribution of network expansion. Third, we analyze the timing of individual investments. Finally, we investigate the robustness of the methods with respect to uncertain input data.

\subsection{The overall welfare performance}
\begin{figure}[ht]
    \centering
    \includegraphics[scale=0.95]{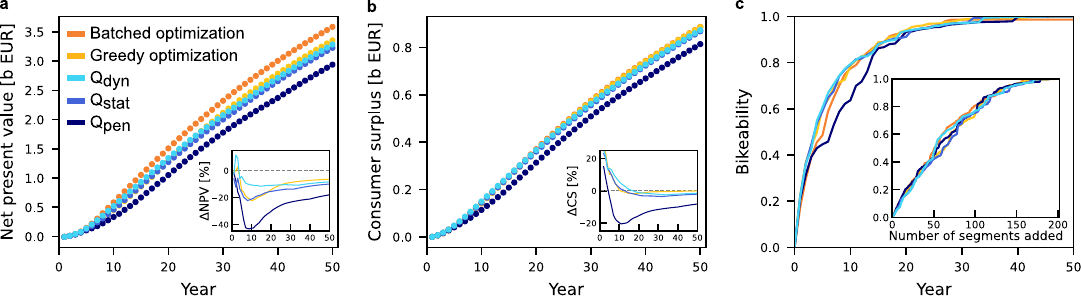}
    \caption{Similar performance of planned networks from dynamic backward percolation and direct optimization. 
    (a) Net present value (NPV), measuring the cumulative societal benefits, including travel time savings, additional cycling demand, and health benefits as well as construction and maintenance cost of the bike paths [Eq.~\eqref{eq:NPV_main}].
    (b) Consumer surplus (CS), measuring the cumulative time saved by all cyclists [see Methods, Eq.~\eqref{eq:CS_methods}]. Insets in panel (a) and (b) show the difference in NPV and CS relative to the batched optimization approach. 
    (c) Bikeability $B$, measuring the relative improvement of travel time and demand between the current and fully build state [Eq.~\eqref{eq:bikeability_main}]. (inset) Bikeability $B$ as a function of the number of segments added. 
    The different approaches achieve similar network performance (panel a-c). Both direct optimization approaches achieve a higher net present value at the end of the 50-year planning horizon as their explicit optimization goal. The networks suggested by the backward percolation approach with the dynamic-demand evaluation function $Q_\mathrm{dyn}$ provide a slightly larger net present value and consumer surplus during the first years of the planning period compared to the batched direct optimization (insets). All approaches achieve a similar bikeability as a function of the number of added segments, suggesting a similar ordering of segment prioritization (inset in panel c). However, the simple evaluation measure $Q_\mathrm{pen}$ for the backward percolation approach falls behind in all measures when evaluated as a function of time, highlighting the importance of maintenance cost and budget constraints.
    }
    \label{fig:FIG4_PerformanceComparison}
\end{figure}

We compare the performance of all network expansion strategies in terms of the net present value, the consumer surplus, and the bikeability index of the networks (Fig.~\ref{fig:FIG4_PerformanceComparison}). Overall, dynamic backward percolation and direct optimization achieve very similar results across the whole planning period. 

As expected, since it more explicitly optimizes for it, direct optimization achieves a higher net present value at the end of the 50-year planning horizon (Fig.~\ref{fig:FIG4_PerformanceComparison}a). The backward percolation with a dynamic-demand evaluation function $Q_\mathrm{dyn}$ initially achieves higher net present value compared to batched optimization in the first 3 years of the planning period (up to \SI{12}{\percent} in year 2, compare Tab.~\ref{tab:TAB1_PlannedNetworks}). This benefit mainly arises due to higher consumer surplus in the first years of the planning period (Fig.~\ref{fig:FIG4_PerformanceComparison}b), but is quickly lost due to higher maintenance cost of the constructed segments. By the end of the planning period, the net present value achieved with the backward percolation approach trails behind by about \SIrange{8}{9}{\percent} compared to batched optimization. Greedy optimization performs significantly worse for the first 25 years (up to \SI{22}{\percent} in year 10). Still, it closes that gap to about \SIrange{6}{7}{\percent} by the end, achieving a slightly higher net present value than backward percolation. With increasing simplification of the segment evaluation, the performance of the backward percolation approach decreases. While the static-demand approximation $Q_\mathrm{stat}$ still achieves comparable results to $Q_\mathrm{dyn}$, the simplified evaluation function $Q_\mathrm{pen}$ falls behind by up to \SI{20}{\percent} after 50 years. 

The network performance regarding the consumer surplus and the bikeability (Fig.~\ref{fig:FIG4_PerformanceComparison}b,c) behaves similarly across all approaches, with differences between direct optimization and percolation approaches becoming smaller for simpler network quality measures. Interestingly, as a function of the number of segments added, the bikeability index is almost identical across all methods. Here, even the simplest evaluation function $Q_\mathrm{pen}$ performs as well or slightly better than the other measures (Fig.~\ref{fig:FIG4_PerformanceComparison}c inset), but loses this advantage when the evaluating the bikeability as a function of time (Fig.~\ref{fig:FIG4_PerformanceComparison}c), emphasizing the importance of including construction and maintenance costs for budget-constrained network expansions.

\begin{table}[ht]
    \centering
    \begin{tabular}{l|r@{\hspace{1.\tabcolsep}}rr@{\hspace{1.\tabcolsep}}rr@{\hspace{1.\tabcolsep}}rr@{\hspace{1.\tabcolsep}}rr}
         &  \multicolumn{2}{c}{Year 3}   &  \multicolumn{2}{c}{Year 10}   &  \multicolumn{2}{c}{Year 30} &  \multicolumn{2}{c}{Year 50} & $G_\mathrm{full}$ \\
         & \multicolumn{1}{c}{DO} & \multicolumn{1}{c}{BP} & \multicolumn{1}{c}{DO} & \multicolumn{1}{c}{BP} & \multicolumn{1}{c}{DO} & \multicolumn{1}{c}{BP} & \multicolumn{1}{c}{DO} & \multicolumn{1}{c}{BP} &  \\ \hline
        Total length built [km] & 338 & 350 & 693 & 705 & 1331 & 1388 & 1331 & 1876 & 1876 \\
        Relative length built [percent] & 18.01 & 18.65 & 36.94 & 37.58 & 70.92 & 73.98 & 70.92 & 100.00 & 100.00 \\
        $\mathrm{CC}_\mathrm{rel}$ [percent] & 6.86 & 6.01 & 20.58 & 20.56 & 54.34 & 57.41 & 54.34 & 100.00 & 100.00\\ \hline
        $\mathrm{NPV}$ [m EUR] & 41.56 & 45.35 & 584.75 & 522.06 & 2321.80 & 2074.91 & 3585.40 & 3284.11 & - \\
        $\mathrm{CS}$ [m EUR] & 14.66 & 17.82 & 141.64 & 149.49 & 575.12 & 562.41 & 886.72 & 873.28 & - \\
        $\mathrm{HB}$ [m EUR] & 43.00 & 41.87 & 498.14 & 427.20 & 1912.15 & 1679.53 & 2929.26 & 2683.81 & - \\
        $\mathrm{CC}$ [m EUR]& 15.18 & 13.28 & 45.50 & 45.46 & 120.14 & 126.92 & 120.14 & 221.10 & 221.10 \\
        $\mathrm{MC}$ [m EUR] & 0.72 & 0.73 & 3.08 & 3.24 & 9.30 & 9.78 & 9.30 & 18.20 & 18.20 \\
    \end{tabular}
    \caption{
    Network costs and performance. 
    Comparison of key observables of the planned networks by batched direct optimization (DO) and by backward percolation (BP) with the dynamic-demand evaluation function $Q_\mathrm{dyn}$. After year 3, both planned networks have similar total length and maintenance costs (MC). Though the construction costs (CC) of BP is \SI{13}{\percent} smaller compared to DO, it achieves \SI{9}{\percent} larger net present value (NPV) due to larger consumer surplus (CS). After year 10, both networks still have a similar total length with almost identical construction costs. Yet, the network planned by BP has \SI{5}{\percent} higher maintenance costs, reducing the available budget for future investments. Also, DO already achieves a \SI{11}{\percent} higher net present value over the BP despite lower consumer surplus due to its advantage in health benefits (HB). After year 30, the batched direct optimization no longer adds additional segments to the network and maintains a net present value approximately \SIrange{8}{10}{\percent} higher than the backward percolation approach. See Supplementary Note 4, Fig.~\ref{fig:FIGS3_Performance_Overview} for the cost and performance for all five approaches.
    }
    \label{tab:TAB1_PlannedNetworks}
\end{table}

\clearpage

\subsection{Geographical distribution}
\begin{figure}[ht]
    \centering
    \includegraphics[scale=0.95]{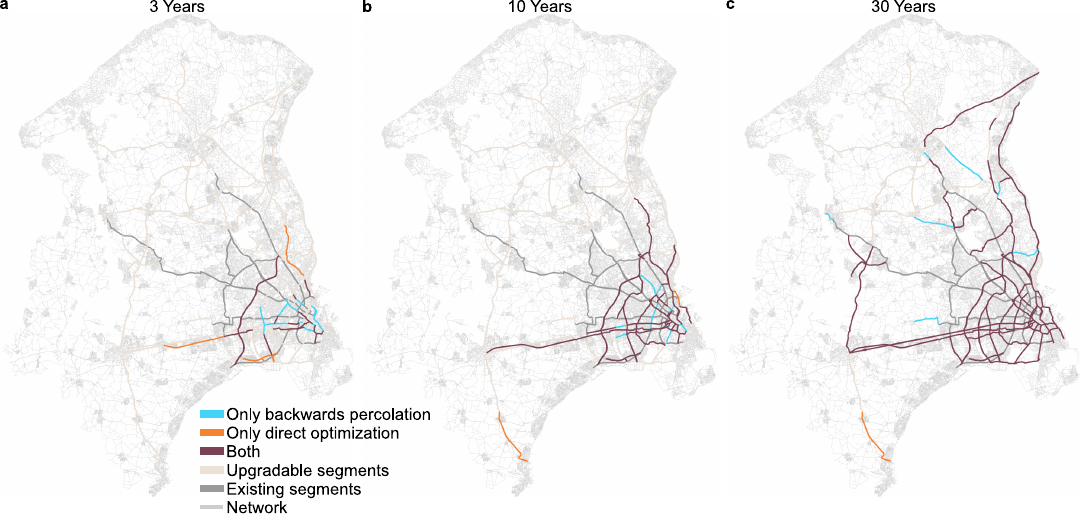}
    \caption{Similar planned networks from dynamic backward percolation and direct optimization. 
    (a-c) Comparison of the planned networks by direct optimization and by backward percolation with the dynamic-demand evaluation function $Q_\mathrm{dyn}$ after (a) 3 years, (b) 10 years, and (c) 30 years (compare Fig.~\ref{fig:FIG2_CopenhagenSetting}a,b, see Tab.~\ref{tab:TAB1_PlannedNetworks} for a quantitative comparison of the costs and benefits of the networks). The planned networks largely coincide (dark red), especially in the urban area of Copenhagen. The direct optimization approach constructs a segment in the south of Copenhagen that remains disconnected from the rest of the already built or existing cycle superhighway network but provides a large net present value benefit (colored or gray, panels b and c).
    }
    \label{fig:FIG5_PlannedNetworks}
\end{figure}

In the following, we compare the geographical distribution of the generated networks between the best-performing direct optimization and backward percolation approaches. Fig.~\ref{fig:FIG5_PlannedNetworks} illustrates the geographical distribution of the proposed expansion after 3, 10 and 30 years (see videos of the network expansion in the Supplementary Material for separate illustrations of the networks of the individual approaches over the 50-year planning period). Initially, in the first three years, both approaches mainly construct relatively short segments in the inner parts of Copenhagen, where demand is highest, and complete a secondary ring around the city (compare Fig.~\ref{fig:FIG5_PlannedNetworks}a). These segments achieve the largest improvement in net present value with relatively low costs. However, batched optimization already constructs some additional radial segments outside the main city area, whereas the backward percolation approach remains focused on the city center. After 10 years, both approaches have built roughly half of the possible segments. The similarity of the suggested networks is high (Fig.~\ref{fig:FIG5_PlannedNetworks}b). Intriguingly, the direct optimization approach constructs a single disconnected segment south of Copenhagen. In contrast, the network suggested by the backward percolation approach is slightly denser in the city center, containing more costly segments. While these segments greatly improve the network, their high maintenance cost restricts the options for the backward percolation approach in the subsequent years, especially in these intermediate planning stages (see Tab.~\ref{tab:TAB1_PlannedNetworks}). This observation explains the better performance of the backward percolation approach in the early years and the loss of performance in the later parts of the planning period because the net available budget is smaller and fewer segments are added annually.
 
After 30 years, both approaches have constructed nearly all planned segments in the urban area of Copenhagen. The planned networks differ by only eight segments. The direct optimization has already reached its final expansion stage. Some newly built segments in the outskirts remain disconnected from the rest of the cycle superhighway network as the direct optimization approach does not add all possible segments (Fig.~\ref{fig:FIG5_PlannedNetworks}c, e.g. the segment in the south, which was built very early on in year 5 and some small segments to the north of Copenhagen).

\clearpage

\subsection{The timing of investments}
\begin{figure}[ht]
    \centering
    \includegraphics[scale=0.95]{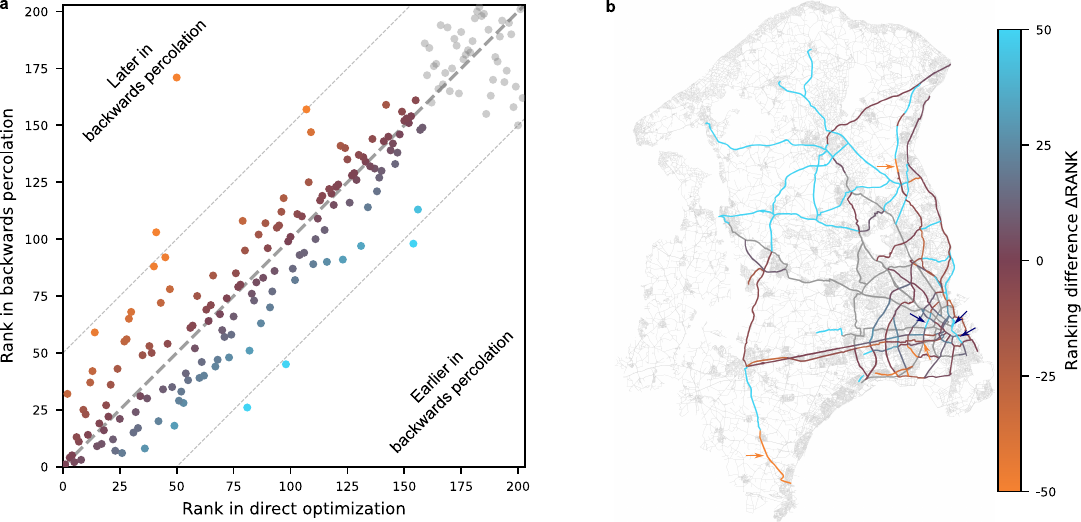}
    \caption{Similar prioritization of cycle superhighway segments in backward percolation and direct optimization. 
    (a) Comparison of the priority rank of each link in the suggested build order in the backward percolation and direct optimization approach. Points on the identity (gray dashed line) indicate identical ranking of segments in the build order, points above (below) the line indicate links that are built later (earlier) in the backward percolation approach. Gray points in the top right represent segments not built by direct optimization. The block structure of the figure suggests that both approaches identify a similar set of least important segments. 
    (b) Map illustrating the difference in raking for each segment (blue: earlier in backward percolation; orange: later in backward percolation, compare panel a). Arrows highlight the seven outliers with a ranking difference $\lvert\Delta\mathrm{STEP}\rvert >= 50$ (outside the dashed light gray lines in panel a). Four of the six outliers are located in the urban area of Copenhagen.
    }
    \label{fig:FIG6_BuildOrderComparison}
\end{figure}

While the planned networks are very similar after 30 years, the timing of the investments up to this point differs. To assess these timing differences between the two planning approaches in more detail, we compare the exact ordering of the segment prioritization (Fig.~\ref{fig:FIG6_BuildOrderComparison}). We evaluate the difference in the ranking of the segments by the two approaches,
\begin{align}
    \Delta \mathrm{RANK}(s) &= \mathrm{RANK}_\mathrm{DO}(s) - \mathrm{RANK}_\mathrm{BP}(s) \,,
\end{align}
where a smaller rank indicates a higher priority (earlier construction). If the difference $\Delta \mathrm{RANK}(s)$ is positive (negative), the segment $s$ is considered more (less) important and built earlier (later) in the backward percolation (BP) approach compared to direct optimization (DO). To be able to compute an explicit ranking of all segments for batched direct optimization, we order the segments constructed in the same year by their rank in the greedy optimization and append segments not built to the end of the ranking.

The results confirm the similarity of the networks generated by both approaches. Over \SI{50}{\percent} of the segments are ranked within a difference $\lvert\Delta\mathrm{RANK}\rvert \leq 9$ and are built within $2$ years of each other. Over \SI{75}{\percent} are ranked within $\lvert\Delta\mathrm{RANK}\rvert \leq 20$ and are built within $4$ years. There are only six outliers ($\lvert\Delta\mathrm{RANK}\rvert >= 50$, see Fig.~\ref{fig:FIG6_BuildOrderComparison}a), constructed more than $5$ years apart by the different planning approaches, one even 29 years. The two outliers built early by direct optimization are located in the perimeter of the planning region (Fig.~\ref{fig:FIG6_BuildOrderComparison}b), notably including the segment in the south that is constructed early and remains disconnected (compare Fig.~\ref{fig:FIG5_PlannedNetworks}b). 

\clearpage

\subsection{Robustness of suggested network extensions}

\begin{figure}[ht]
    \centering
    \includegraphics[scale=0.95]{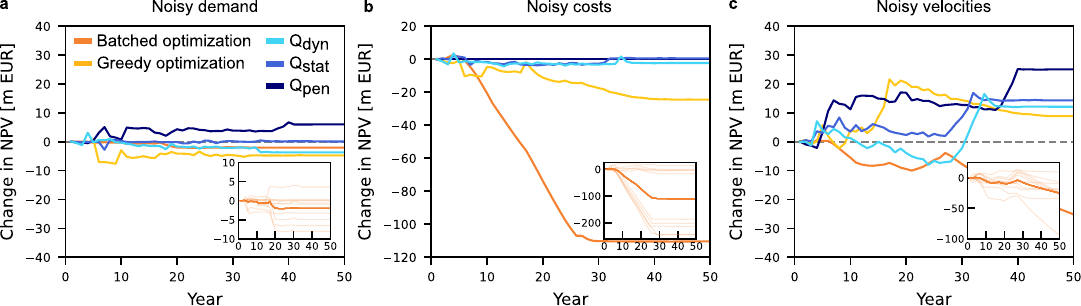}
    \caption{Robustness of suggested cycle superhighway extensions decreases with model complexity. 
    (a-c) Change of the net present value of the networks planned with noisy data compared to networks planned with the exact parameters for (a) noisy demand, (b) noisy costs, and (c) noisy velocities. The insets show the result for the single runs and the mean, exemplary for the batched optimization. While the simplest importance measure $Q_\mathrm{pen}$ of backward percolation results in the smallest net present value (compare Fig.~\ref{fig:FIG4_PerformanceComparison}a), it is overall the most robust to variations of the input data. With increasing model complexity, the change in net present value increases on average, indicating an effect similar to overfitting of the optimized network expansion order to the input data.
    }
    \label{fig:FIG7_Robustness}
\end{figure}

Any network planning application is subject to uncertainty, whether due to changing conditions that cannot be captured in the model or inaccurate and noisy data underlying the planning decisions. Here, we test the robustness of the cycle superhighway extensions with respect to noise in three different parts of the input data by varying the original values by up to $\pm \SI{20}{\percent}$: (i) We consider noisy demand by adding multiplicative noise with a uniformly random factor $\lambda \in [-0.2,0.2]$ to the demand from each zone (compare Fig.~\ref{fig:FIG2_CopenhagenSetting}), adjusting the trip demand for an origin-destination-pair $(o,d)$ by the average of the noise factor of both the origin and destination zone. (ii) We consider noisy costs by adding multiplicative noise with a uniformly random factor $\lambda \in [-0.2,0.2]$ to each cycle superhighway segment's construction and maintenance cost independently. (iii) We consider noisy velocity values for the cyclists by adding multiplicative noise with a uniformly random factor $\lambda \in [-0.2,0.2]$ to each speed-cyclist-infrastructure velocity, ensuring that the relative ordering of the speeds inside categories remains the same (compare Methods, Tab.~\ref{tab:TAB2_CyclistTypes}).
The aim is to test the robustness of the cycle superhighway extension with respect to current uncertain knowledge and misestimation of the model parameters.
We compute network extensions for ten samples of noisy input data for each of the three variations and compare the resulting average deviations of the net present value (Fig.~\ref{fig:FIG7_Robustness}), consumer surplus, and bikeability (Supplementary Note 5, Fig.~\ref{fig:FIGS4_Robustness_Metrics_Mean}). Results of all individual realizations are presented in Supplementary Note 5 Figs.~\ref{fig:FIGS5_Robustness_Ranks} to \ref{fig:FIGS10_Robustness_BaS_all}.

To evaluate the robustness of the different methods, we compare the performance of the planned cycle superhighway extensions with noisy input data to the original planning with non-noisy parameters of the respective method (Fig.~\ref{fig:FIG7_Robustness}). Inaccurate demand data has almost no effect on the resulting network quality for all five methods. This is likely due to the large number of origin-destination pairs averaging out the effect of the noise. 

Uncertain segment costs have a larger influence on the planned network performance, especially for the direct optimization approaches. In contrast, the backward percolation with the simplest quality evaluation function $Q_\mathrm{pen}$ is not affected at all, since it does not explicitly consider segment costs. The batched optimization is highly affected and loses up to 240 m EUR corresponding to $2/3$ of its additional net present value compared to the backward percolation models (Fig.~\ref{fig:FIG7_Robustness}b inset).

Noisy velocities have a strong impact across all approaches, likely because the velocities play a key role in routing the cyclists and especially since a binary shortest path route choice model is used. Here, even planning based on the simplest importance measure $Q_\mathrm{pen}$ is strongly influenced, as it heavily relies on the ratio of the velocities to define the importance of a cycle superhighway segment. Intriguingly, the quality of the planned networks on average improves slightly for all approaches except batched optimization, which already achieves a higher net present value than the other approaches.

For a detailed explanation of the robustness dynamics we have to differentiate between two areas (i) general, approach-based differences/dynamics and (ii) specific, data- and network-dependent differences.

In general, the lower robustness of the batched approach can be attributed to its near-optimal performance with the base non-noisy input data. Any change in the proposed ordering is more likely to (strongly) reduce the benefits. In contrast, for the simpler models, including the greedy optimization, changes in the ordering are more likely to also result in better performance. At the same time, the network evaluation measures used in these simpler models also depend on fewer parameters and combine fewer terms, resulting in overall smaller fluctuations.

For this specific network, a large change in the net present value is caused by a single (long) segment of the cycle superhighway (visible for example in the qualitative change of the trajectories in the inset in Fig.~\ref{fig:FIG7_Robustness}b). Delaying the construction of this segment causes a significant decrease in the net present value of the resulting network. However, due to the size and cost of the segment, estimates of the benefits and its impact on routing decisions are unstable. Consequently, the evaluation of the segment changes drastically and it is often built at different times, especially with noisy costs or velocities.

Overall, variations in the build order and correlation of the ranking of the segment importance (see Supplementary Note 5, Fig.~\ref{fig:FIGS5_Robustness_Ranks}) confirm the qualitative observations: more complex optimization algorithms are typically more susceptible to inaccurate input data. These results highlight the benefits of simplified, abstract models when data is known to be uncertain.


\section{Discussion}

Planning efficient network extensions is a highly complex problem across all applications due to multiple, interacting, and often opposing constraints and requirements \cite{Aldous2019_OptimalGeometryTransportation, Gastner2006_OptimalDesignSpatial}. Here, we have directly compared two complementary network planning approaches for the design of cycle superhighway extensions with the same underlying network and demand data. While direct optimization approaches achieve a higher net present value, our results show that abstract conceptual planning approaches may achieve comparable performance with simpler and often more robust evaluation functions.

Both approaches have advantages and disadvantages, but complement each other in their applicability for different use cases. The backward percolation approach produces more reliable solutions to the problem in case of uncertain and absent input data. This makes it useful as a first step in the network planning phase. However, the backward planning makes it impossible to evaluate properties depending explicitly on the construction time of individual infrastructure elements, such as the total maintenance costs until the end of the planning period. In contrast, the direct optimization approaches construct the network from the ground up with explicit reference to the budgetary constraints \cite{Paulsen2023_SocietallyOptimalExpansion, Paulsen2024_WelfareOptimalExpansionInducedDemand}. This method achieves better performance for the explicit optimization goal and is likely more suitable for constructing accurate expansion plans later in the planning process when all parameters and conditions are known sufficiently accurately. However, simplifications to the optimization problem, like the linearization of travel time improvements (see Methods), seem to result in more disconnected bike paths in intermediate network stages compared to other approaches that focus on conceptual measures and put more emphasis structural properties of the networks. While we have only tested the robustness of the suggested network extensions to variations in the input data, planning problems often suffer from increasing uncertainty of external conditions and model parameters during the planning period. A valuable extension of this research would be exploring how such progressively increasing uncertainty affects solution strategies, especially given the time horizons considered.
 Additionally, a route choice model that does not solely rely on travel speeds would be a beneficial extension. Yet, such a route choice model still needs to be computationally efficient when used with a rapidly changing network, excluding standard logit models that rely on predefined choice sets.

Neither approach fully automates the planning process, as these processes often exhibit multiple stages and revisions. Furthermore, the prioritization and actual implementation of extension measures requires much more information than can be represented in any algorithmic model. Not only is the participation of residents and businesses necessary in the planning process \cite {Zhao2018_bicycle, Brabham2009_crowdsourcing}, but as other modes of transport are often impacted by the construction of new bike paths, a comprehensive urban planning perspective has to include a multi-modal evaluation \cite{Ballo2023_EbikeCity}. Additional robustness analyses may provide more detailed insights on the variability of the planned extensions with respect to such model extensions, for example more detailed route choice models or additional optimization criteria representing costs and benefits for other road users and stakeholders.

Regardless of the potential extensions, the suggested networks may serve as a basis for planning and supporting an efficient prioritization of planned network extensions in such a multi-stage process. After the identification of possible extensions, which are politically and technically feasible, our model could help to rank and prioritize these planned extensions. With this ranking, city planners could start detailed planning, including community feedback and add-on effects on other road users. Our algorithmic planning approaches may also help to identify critical extensions that, if proven unfeasible or too costly in these stages, would result in large changes of the suggested network expansion. This might help to focus efforts on important segments that require detailed evaluation early in the planning process to identify robust long-term expansion strategies.

In general, both network planning approaches may (with some adaptions) also be applied to general infrastructure networks \cite{Barthelemy2011_SpatialNetworks}, including multi-modal transport infrastructure \cite{Alessandretti2023_multimodal, Olafsson2016_cycling, Caggiani2020_equality}, charging infrastructure for electric vehicles and power grids \cite{Marszal2021_phase, Schafer2022_understanding}, or planning resilient infrastructure for climate change \cite{Turner2023_shade, Krayenhoff2021_cooling, Ramyar2021_adapting}. While a wide range of other methodologies, such as reinforcement learning often used in approximate dynamic programming, have been applied to solve network planning problems in these different contexts, the scalability of many methods to large network contexts emerging in many of these applications remains an open question.

Overall, our results show that conceptual network design approaches from network science may compete with direct optimization methods and complement existing approaches. Together, both approaches may help design more efficient bike path networks to support cycling and the transition to more sustainable urban mobility.

\clearpage

\section{Methods}

\subsection*{Copenhagen cycle superhighway model}

\paragraph{Network}
For the numerical implementation, we represent the street, bike path, and cycle superhighway network as a directed graph $G = (V,E)$. The vertices represent intersections as well as intermediate points along single streets, bike paths, or cycle superhighways. The edges $E$ represent the directed connections along the streets, bike paths, or cycle superhighways. Each directed edge $e \in E$ is assigned a physical length $l_e$ and one of three infrastructure categories (street, bike path, cycle superhighway), determining the travel speed of cyclists on the edge (see below).

We distinguish between the base network $G_\mathrm{base} = (V_\mathrm{base}, E_\mathrm{base})$ without any of the proposed cycle superhighway upgrades (Fig.~\ref{fig:FIG2_CopenhagenSetting}a) and the planned upgrades $G_\mathrm{build}$ (Fig.~\ref{fig:FIG2_CopenhagenSetting}b). Their combination constitutes the fully upgraded network, $G_\mathrm{full} = G_\mathrm{base} \cup G_\mathrm{build}$. Over the course of the optimization, we add a total of $S = 202$ cycle superhighway segments $E_s$, $s \in \{1,2,\ldots,S\}$, each consisting of multiple individual edges, $E_s = \{e_1(s), e_2(s), \ldots \}$. These edges may represent upgrades to existing edges in the network $G_\mathrm{base}$ or entirely new connections. When a cycle superhighway segment is added to the graph $G$, we add its edges and vertices to the network (if the edges did not exist in the network yet) or update the infrastructure types of the edges included in the segment to cycle superhighway (if the edges were already part of the network as streets or normal bike path). 

The full network $G_\mathrm{full}$ (compare Fig.~\ref{fig:FIG2_CopenhagenSetting}a,b) has $\lvert V\rvert = 191448$ vertices and $\lvert E\rvert = 453433$ directed edges with a total length of approximately \SI{33678}{\km}, double-counting directed edges in both directions along the same path. A large part of the network is already covered by the normal bike path network, accounting for \SI{9894}{\km} (\SI{29.4}{\percent}) of the length of the network. The existing cycle superhighway network makes up \SI{460}{\km} (\SI{1.4}{\percent}) with the proposed upgrades adding \SI{1876}{\km} (\SI{5.6}{\percent}). Of the proposed segments, \SI{1666}{\km} are upgrades based on existing links within the base network and \SI{210}{\km} are completely new infrastructure.

The construction cost\cite{Incentive2018} of the proposed cycle superhighway segments range between \SI{5468.73}{\myeuro\per\km} and \SI{1245096.76}{\myeuro\per\km}, with an average of \SI{138978.02}{\myeuro\per\km} (median \SI{96972.49}{\myeuro\per\km}).
Their annual maintenance cost\cite{Incentive2018} range between \SI{636.41}{\myeuro\per\km} and \SI{98621.53}{\myeuro\per\km}, with an average of \SI{11484.49}{\myeuro\per\km} (median \SI{7441.89}{\myeuro\per\km}).

\paragraph{Demand}
The demand model consists of $52\,808$ origin-destination pairs between $258$ starting and end points (compare Fig.~\ref{fig:FIG2_CopenhagenSetting}c), representing the annual number of cyclists \cite{Hallberg2021_CycleSuperHighway}. The demand for each origin-destination pair is further split into $9$ different cyclist types, reflecting the fitness or general speed of the cyclists (slow, medium, fast) and the type of bicycle (regular bicycle, e-bike and speed pedelec) \cite{Hallberg2021_CycleSuperHighway}. Normal bikes make up \SI{95}{\percent} of the trips, e-bikes \SI{4.5}{\percent}, and speed pedelecs account for the remaining \SI{0.5}{\percent}. We assume that \SI{50}{\percent} of cyclists belong to the medium speed type and \SI{25}{\percent} to the slow and fast type, respectively. Each cyclist type is assigned a velocity $v$ for each of the three infrastructure categories of edges in the network (street, bike path, cycle superhighway), see Tab.~\ref{tab:TAB2_CyclistTypes}, summarizing the actual travel time as well as perceived comfort and safety of travel along the edge.

We model induced demand as a function of the travel time $\tau(\omega,G)$ of cyclists for a trip-cyclist-type combination $\omega$ in the network $G$ with a logit model \cite{Train2009_DiscreteChoice}. The travel demand $n(\omega, G)$ changes as
\begin{align}
    n(\omega,G) = n(\omega) \, P(\omega, G) = n(\omega) \, \frac{e^{-\beta\,\tau(\omega,G)}}{e^{-\beta\,\tau(\omega,G)} + e^{-\beta\,\tau_\mathrm{other}(\omega)}}
\end{align}
where $n(\omega)$ denotes the total travel demand for the trip $\omega$ and $\tau_\mathrm{other}(\omega)$ the constant travel time for the trip $\omega$ with other modes of transport. The parameter $\beta \approx \SI[per-mode=power]{-0.0518}{\per\hour}$ is the parameter that describes the demand sensitivity. This parameter is calibrated to render elasticities similar to the ones used by Hallberg \emph{et~al.} \cite{Hallberg2021_CycleSuperHighway} and Paulsen \& Rich \cite{Paulsen2024_WelfareOptimalExpansionInducedDemand}.

For the evaluation of the net present value, we include population growth. The trip demand in a network $G$ at time $t$ is given by
\begin{align}
    n(\omega,G,t) = \gamma(t) \, n(\omega,G) = \gamma(t) \, \frac{n(\omega,G_\mathrm{base})}{P(\omega,G_\mathrm{base})}\, \frac{e^{-\beta\,\tau(\omega,G)}}{e^{-\beta\,\tau(\omega,G)} + e^{-\beta\,\tau_\mathrm{other}(\omega)}}
\end{align}
with the prefactor $\gamma(t)$ capturing overall population growth, with $\gamma(0) = 1$ and increasing by approximately $\SI{0.2}{\percent}$ per year for a total increase of $\SI{7}{\percent}$ by the end of the 50-year planning period \cite{StatisticsDenmarkGrowth}. Here we write the total demand $n(\omega) = \frac{n(\omega,G_\mathrm{base})}{P(\omega,G_\mathrm{base})}$ as a function of the demand $n(\omega,G_\mathrm{base})$ and the fraction $P(\omega,G_\mathrm{base})$ of cyclists in the base network.

\clearpage

\paragraph{Route choice model}
Modelling cyclist route choices in a changing network environment is in itself a complex problem \cite{Lukawska2024_quantitative}. 
Cyclist routing is largely unaffected by congestion, as low space requirements and cyclist volumes keep travel times stable, with uncertainties mainly from intersection delays\cite{Paulsen2019_congestion}. We therefore  consider each trip individually and adopt a simplified shortest path choice model. This has the added benefit of avoiding complication and bias introduced by explicit generation of choice sets as the network changes over time.
Hence, cyclists travel through the network following the shortest path with respect to their travel time, determined by their bike type and speed on the different types of network edges as explained above (Tab.~\ref{tab:TAB2_CyclistTypes}, derived from empirical speed measurement from cyclists in Denmark on different infrastructure types \cite{Hallberg2021_CycleSuperHighway}). In addition to the travel time along the network edges, we add time penalties for complex intersections. These penalties are the same for all cyclist types, adding \SI{30}{\second} for large traffic light-controlled intersections, \SI{5}{\second} for roundabouts, and \SI{0}{\second} for the remaining ones (including, for example, intermediate vertices along paths, intersections between small residential streets, and on- and off-ramps of the cycle superhighway network) \cite{Hallberg2021_CycleSuperHighway}.

\begin{table}[ht]
    \centering
    \begin{tabular}{cc|ccc}
        Bicycle Type    & Speed Type    & \multicolumn{3}{c}{Infrastructure Type} \\
                        &               & Street & Bike Path & Cycle Superhighway \vspace{2mm}\\
                        \hline\\[-3mm]
                        & Slow          & 13.6   & 15.1      & 16.6 \\
        Regular Bicycle & Medium        & 16.3   & 17.8      & 19.3 \\
                        & Fast          & 19.1   & 20.8      & 22.5 \vspace{1mm}\\
                        \hline\\[-3mm]
                        & Slow          & 15.6   & 17.1      & 18.6 \\
        E-bike          & Medium        & 18.3   & 19.8      & 21.3 \\
                        & Fast          & 21.1   & 22.8      & 24.5 \vspace{1mm}\\
                        \hline\\[-3mm]
                        & Slow          & 22.6   & 24.1      & 25.6 \\
        Speed Pedelec   & Medium        & 25.3   & 26.8      & 28.3 \\
                        & Fast          & 27.3   & 29.8      & 31.5 \\
    \end{tabular}
    \caption{Travel speeds in \si{\km\per\hour} of cyclist types based on Hallberg \emph{et~al.}\cite{Hallberg2021_CycleSuperHighway}.}
    \label{tab:TAB2_CyclistTypes}
\end{table}

\clearpage

\subsection*{Net present value of the cycle superhighway network}
We measure the societal benefits of the cycle superhighway network extensions in terms of the net present value [NPV, Eq.~\eqref{eq:NPV_main}] of the network $G$ \cite{Paulsen2023_SocietallyOptimalExpansion, Paulsen2024_WelfareOptimalExpansionInducedDemand}. The net present value includes five components:

(i) The annual consumer surplus (CS),
\begin{align}\label{eq:CS_methods}
    \mathrm{CS}(G,t) = \sum_\omega \zeta(\omega)\,\frac{n(\omega,G_\mathrm{base}) + n(\omega,G,t)}{2} \, \left[\tau(\omega,G_\mathrm{base}) - \tau(\omega,G)\right] \,,
\end{align}
in year $t$ measures the travel time savings $\tau(\omega, G_\mathrm{base}) - \tau(\omega,G)$ of cyclists across all trip-cyclist-type combinations $\omega$ compared to the base network $G_\mathrm{base}$, employing the rule-of-half approximation to account for induced demand \cite{Kidokoro2004}. Here, $\tau(\omega, G)$ and $\tau(\omega, G_\mathrm{base})$ denote the travel time of cyclists and $n(\omega,G,t)$ and $n(\omega, G_\mathrm{base})$ denote the cycling demand in the current and base network, respectively. The factor $\zeta$ represents the value of time, converting the travel time savings into a monetary value.

(ii) The annual health benefits (HB),
\begin{align}
    \mathrm{HB}(G, t) &= \sum_\omega \xi(\omega) \left[n(\omega, G, t) \, \ell(\omega, G) -  n(\omega, G_\mathrm{base}) \, \ell(\omega, G_\mathrm{base}) \right] \,,
\end{align}
in year $t$ capture the value of increased distance cycled over all trip-cyclist-type combinations $\omega$, mainly due to induced demand. Here, $\ell(\omega, G)$ and $\ell(\omega, G_\mathrm{base})$ denote the physical trip distances in the current and base network, respectively. The factor $\xi$ again represents the conversion factor of the health benefits to monetary value.

(iii) The total construction costs (CC),
\begin{align}
    \mathrm{CC}(G) &= \sum_{s \in S_{G}} \mathrm{cc}(s) \,,
\end{align}
denote the total financial investment in the cycle superhighway extensions for all built segments $S_{G}$ in the graph $G$, where $\mathrm{cc}(s)$ are the construction cost for the individual segment $s$.

(iv) The annual maintenance costs (MC),
\begin{align}
    \mathrm{MC}(G) &= \sum_{s \in S_{G}} \mathrm{mc}(s) \,,
\end{align}
denote the required financial investment for maintaining the current cycle superhighway extensions for all built segments $S_{G}$ in the graph $G$, where $\mathrm{mc}(s)$ are the annual maintenance cost for the individual segment $s$.

(v) Finally, the total scrap value (SV),
\begin{align}
    \mathrm{SV}(t, G) &= \kappa(t) \sum_{s \in S_{G}} \,\mathrm{cc}(s) \,,
\end{align}
denotes the remaining value of the construction costs of the cycle superhighway extensions at time $t$. The factor $\kappa(t)$ captures the devaluation as a function of time.

The net present value of a sequence of networks $\{G(t)\}$ up to year $t$ of the planning stage (including the newly built cycle superhighway extensions) is then given by the sum of the annual components up to year $t$ plus the one-time contributions of construction cost and scrap value. Importantly, the net present value contribution of each year is weighted with the devaluation factor $\kappa(t)$. In total, we have
\begin{align}
     &\phantom{=} \mathrm{NPV}(t, \{G(t)\}) \nonumber \\
     &= \left[ \sum_{t' = 2}^{t} \kappa(t') \, \mathrm{CS}(G(t'), t') \right]
      + \left[ \sum_{t' = 2}^{t} \kappa(t') \, \mathrm{HB}(G(t'), t') \right]
      - \Bigg[ \mathrm{CC}(G(t)) \Bigg]
      - \left[ \sum_{t' = 2}^{t} \kappa(t') \, \mathrm{MC}(G(t'-1)) \right]
      + \Bigg[ \mathrm{SV}(t, G(t)) \Bigg] \nonumber\\
     &= \left[ \sum_{t' = 2}^{t} \kappa(t') \, \sum_\omega \zeta(\omega)\,\frac{n(\omega, G_\mathrm{base}) + n(\omega, G(t'), t')}{2} \, \Big(\tau(\omega, G_\mathrm{base}) - \tau(\omega, G(t'))\Big) \right]  \nonumber\\
     &\phantom{=} + \left[ \sum_{t' = 2}^{t} \kappa(t') \sum_\omega \xi(\omega) \Big(n(\omega, G(t'), t') \, \ell(\omega, G(t')) -  n_\omega(G_\mathrm{base}) \, \ell(\omega, G_\mathrm{base}) \Big) \right] \nonumber\\
     &\phantom{=} - \left[ \sum_{s \in S_{G(t)}} \mathrm{cc}(s) \right] 
      - \left[ \sum_{t' = 2}^{t} \kappa(t) \sum_{s \in S_{G(t'-1)}} \mathrm{mc}(s) \right]
      + \left[ \kappa(T) \sum_{s \in S_{G(t)}} \mathrm{cc}(s) \right] \,, 
\end{align}
where the summation starts in the second year, after the first planned extensions have been constructed in year $t=1$, where only construction and scrap value contribute to the net present value.

\clearpage

\subsection*{Direct optimization of the cycle superhighway network}
\paragraph{Batched optimization}
The direct optimization model from Paulsen \& Rich (2024)\cite{Paulsen2024_WelfareOptimalExpansionInducedDemand} iteratively finds the best cycle superhighway segments to be built by maximizing the predicted net present value at the end of the 50-year planning period.
The model incorporates expected changes in consumer surplus and health benefits, including those driven by increased demand, to identify the optimal combination of segments to be constructed each year. The objective function $\Delta\,\tilde{\mathrm{NPV}}$ approximating the gain in net present value, defined for all segments not yet included in the graph $G$ at a given period $t$, is expressed as 
\begin{align}
    \Delta\,\tilde{\mathrm{NPV}}(s,G,t) 
    &= \left[\sum_{t' = t+1}^{T} \kappa(t') \right] \, \Delta\tilde{CS}(s,G,t)\, + \left[\sum_{t' = t+1}^{T} \kappa(t') \right]\Delta\tilde{HB}(s,G,t)    -  \kappa(t) \,\mathrm{cc}(s) 
      - \left[\sum_{t' = t+1}^{T} \kappa(t')\right] \, \mathrm{mc}(s) \,.
\end{align}

Here, $\tilde{CS}$ represents the linearized estimates of raw consumer surplus, approximated using the travel times of the network $G$ before the current decision step and the fully expanded network, $G_\mathrm{full}$:
\begin{align}
 \Delta \tilde{\mathrm{CS}}(s,G,t) = \sum_\omega \zeta(\omega)\,\frac{n(\omega,G_\mathrm{base}) + \tilde{n}(\omega,s,G,t)}{2} \,  \left[\tau(\omega,G) - \tilde{\tau}(\omega,s,G) \right]\,, \label{eq:CSTilde}
\end{align}
with $\tilde{\tau}(\omega,s,G)$ defined as
\begin{align}
         \tilde{\tau}(\omega,s,G) 
    &= \tau(\omega,G) - \Big(\tau(\omega,G) - \tau(\omega,G_\mathrm{full})\Big) \frac{d\left(\omega,s,G_\mathrm{full}\right)}{\sum_{s' \notin G}d\left(\omega,s',G_\mathrm{full}\right)} \,, \label{eq:TauTilde}
\end{align}
and $\tilde{n}(\omega,s,G,t) $ as
\begin{align}
         \tilde{n}(\omega,s,G,t) 
    &= \gamma(t)\,n(\omega,\tilde{\tau}(\omega, s,G))\,. \label{eq:NTilde}
\end{align}
The linearized estimations of health benefits $\Delta\tilde{HB}$ are similarly defined as 
\begin{align}
         \Delta \tilde{\mathrm{HB}}(\omega,s,G) 
    &=   \sum_\omega \xi(\omega) \left[\tilde{n}(\omega,s, G, t) \, \tilde{\ell}(\omega,s,G) -  n(\omega, G_\mathrm{base}) \, \ell(\omega, G_\mathrm{base}) \right] \,, \label{eq:HBTilde}
\end{align}
with
\begin{align}
         \tilde{\ell}(\omega,s,G) 
    &=   \ell(\omega,G) - \left(\ell(\omega, G) - \ell(\omega, G_\mathrm{full})\right)\frac{d\left(\omega,s,G_\mathrm{full}\right)}{\sum_{s'\notin G}d\left(\omega,s',G_\mathrm{full}\right)}\,. \label{eq:EllTilde}
\end{align}

The concept behind the linearization \cite{Paulsen2023_SocietallyOptimalExpansion} is that, for each $\omega$, the changes in travel times and distances resulting from upgrading the entire network can be attributed to the segments utilized in the full network, $G_\mathrm{full}$. The individual weighting of each segments $s$ (as represented by the fractions in Eq.~\eqref{eq:TauTilde}, Eq.~\eqref{eq:NTilde}, and Eq.~\eqref{eq:EllTilde}) is proportional to the distance traveled on $s$ in the full network $G_\mathrm{full}$. If the denominator is zero, the numerator will also be zero. In this case, the fraction should be treated as a zero and thus does not influence $\tilde{\tau}$, $\tilde{n}$ or $\tilde{\ell}$.  

Based on the predicted change $\Delta\,\tilde{\mathrm{NPV}}(s,G,t)$ in the net present value for each segment $s$, we formulate a binary integer program over the set of non-selected segments. The problem is solved using a standard solver\cite{lpSolve} to find the optimal set $S_t = \left\{s^*_{t,1}, s^*_{t,2}, \ldots\right\}$ to be built in year $t$. This set of segments maximizes the expected change in net present value subject to the budget constraint in the current year $t$, 
\begin{align}
    \sum_{s \,\in\, S(t)} \mathrm{cc}(s) \le \sum_{t' = 1}^{t} b(t') \;- \sum_{s \,\in\, S_{< t}} \mathrm{cc}(s) - \sum_{t'=2}^t \, \sum_{s \,\in\, S_{< t'}} \mathrm{mc}(s)
    ,\label{eq:BudgetConstraintDTU}
\end{align}
where $b$ denotes the (constant) annual budget and $S_{< t} = \bigcup_{t' = 1}^{t-1} S_{t'}$ denotes the set of cycle superhighway segments constructed up to but not including year $t$. The second sum on the right-hand-side thus describes the total construction cost of all segments built until and including year $t-1$, the third sum describes the cumulative maintenance cost for all constructed segments, e.g.~in year $t$ we pay the maintenance cost for all cycle superhighway segments constructed in year $t-1$ and before.

After each year $t$, we add the set $S_t$ to $G$ and recompute the estimated net present value changes for each remaining segment, update the budget constraints and select the next set of segments to be constructed until the expected change in net present value is negative for all remaining segments. The update from year to year requires recomputing the routes for each $\omega$ in order to update the actual and estimated travel times and distances of the shortest paths for each $\omega$.

\newpage

\paragraph{Greedy optimization}
We also analyze a simpler version of the optimization algorithm from Paulsen \& Rich (2023) \cite{Paulsen2023_SocietallyOptimalExpansion}. This version differs from the more advanced approach \cite{Paulsen2024_WelfareOptimalExpansionInducedDemand} described above by selecting segments greedily based on an estimated net present value rate per construction cost $\Delta \tilde{R}$  under the assumption of constant demand (comparable to the advanced evaluation functions for the backward percolation approach, see below),
\begin{align}
    \Delta\,\tilde{\mathrm{R}}(s,t) 
    &= \frac{ \left[\sum_{t' = t+1}^{T} \kappa(t') \right] \, \Delta\tilde{CS}(s,G_\mathrm{base},0)\,    -  \kappa(t) \,\mathrm{cc}(s) 
      - \left[\sum_{t' = t+1}^{T} \kappa(t')\right] \, \mathrm{mc}(s) }{  \kappa(t)\,\mathrm{cc}(s)}\,.
\end{align}
This greedy approach does not consider optimal bundling of segments within a period and entirely excludes health benefits, as they are considered negligible under the assumption of constant demand.

As before, construction costs contribute immediately, while consumer surplus and maintenance costs contribute only in the years following construction ($t' \geq t+1$). It is assumed that the contribution from each segment $s$ can be isolated, independent of other segments. In contrast to the batched optimization approach introduced above, the simplified consumer surplus estimates are computed based only on the initial and final network state $G_\mathrm{base}$ and $G_\mathrm{full}$ without recomputing the routes or travel times of cyclists. The approximated consumer surplus associated with segment $s$ is thus a simplified case of Eq.~\eqref{eq:CSTilde},
\begin{align}
  \Delta \tilde{\mathrm{CS}}(s,G_\mathrm{base},0) = \sum_\omega n(\omega,G_\mathrm{base}) \,  \left[\tau(\omega,G_\mathrm{base}) - \tilde{\tau}(\omega,s,G_\mathrm{base}) \right]\,. \label{eq:CSTildeSimple}
\end{align}

Using the predicted change $\Delta R(s,t)$ in net present value per construction cost for each segment $s$, we greedily select the set $S_t = \left\{s^*_{t,1}, s^*_{t,2}, \ldots\right\}$ of segments to be constructed in year $t$. This set maximizes the expected change in net present value, satisfying $\Delta \tilde{\mathrm{NPV}}(s^*_{t,1}) \geq \Delta \tilde{\mathrm{NPV}}(s^*_{t,2}) \geq \ldots$, while adhering to the same budget constraint as specified in the advanced model (Eq.~\eqref{eq:BudgetConstraintDTU}).
After each year, we recompute the estimates of the net present value rates and select the next set of segments to be built until all cycle superhighway segments have been added to the network (after 47 years). The update from year to year is much faster compared to the advanced method, as only the discount factors $\kappa$ depend on $t$. However, evaluation of the network quality requires additional post-processing and computation of the routes and travel times of all cyclists.


\subsection*{Dynamic backward percolation approach}
The backward percolation algorithm starts from the full network and iteratively removes the least important cycle superhighway segment from the network. The general procedure is as follows:

\subsubsection*{Algorithm}
{\setlist{nolistsep}
\begin{enumerate}[noitemsep, label=\arabic*.]
\item Set up initial network state $G \leftarrow G_\mathrm{full}$
\begin{enumerate}[noitemsep,label*=\alph*]
     \item Read in base network $G_\mathrm{base}$
     \item Add all buildable cycle superhighway segments $s \in S$
     \item Calculate the routes of all trips
\end{enumerate}
\item Main loop, iterate until all buildable cycle superhighway segments have been removed and $G = G_\mathrm{base}$.
\begin{enumerate}[noitemsep,label*= \alph*]
    \item Calculate the evaluation function $Q(s)$ for each remaining segment $s \subset G$.
    \item Select the least important remaining segment $s^*$ with $Q(s^*) \le Q(s)$ for all $s \subset G$.
    \item Remove the segment from the network, $G \leftarrow G \setminus s^*$.
    \item Recalculate the routes for all affected trips
\end{enumerate}
 \item Algorithm finished.
\end{enumerate}}

\paragraph{Evaluation criteria}
We use three different evaluation criteria $Q$ to determine the least important cycle superhighway segment to be removed in each step of the algorithm:
\begin{enumerate}
    \item [(i)] The normalized travel time penalty $Q_\mathrm{pen}$ introduced in the original backward percolation model \cite{Steinacker2022_DemandDrivenDesign},
    \begin{align}
        Q_\mathrm{pen}(s, G) &= \frac{\sum_\omega \sum_{e \in s} n_e(\omega, G) \, l_e \, c_e(\omega)}{\sum_{e \in s} l_e} \,.
    \end{align}
    The penalty $c_e(\omega) = \frac{v_\mathrm{csh}(\omega)}{v_e(\omega)} > 1$ describes the relative change in travel time for cyclists when the cycle superhighway is removed, assuming cyclists do not change their routes. Here $v_\mathrm{csh}(\omega)$ denotes the speed of cyclists traveling in the cycle superhighway network, while $v_e(\omega)$ denotes the speed of cyclists traveling in the original base network (compare Tab.~\ref{tab:TAB2_CyclistTypes}). This measure is proportional to the total travel time for all cyclists $n_e(\omega, G)$ using edges $e$ of the segment $s$. For newly built edges, we take the penalty $c$ as the penalty of a street edge to avoid infinite importance of segments.
    \item [(ii)] The relative change in consumer surplus $Q_\mathrm{stat}$ in consumer surplus,
    \begin{align}
        Q_\mathrm{stat}(s, G) &= \frac{\sum_\omega \sum_{e \in s} \zeta(\omega) \frac{n(\omega, G_\mathrm{base}) + n(\omega, G)}{2} 
        \Delta \tau_e(\omega)}{\mathrm{CC}(s)} \\
        &\approx \frac{\Delta \mathrm{CS}(s,G)}{\mathrm{CC}(s)} \,,  \nonumber
    \end{align}
    representing the relative change in net present value compared to the required investment costs due to travel time reduction, assuming that demand and route choices do not change when the segment is removed. Here, $\Delta \tau_e(\omega) > 0$ describes the change in travel time when the cycle superhighway segment is removed.
    \item [(iii)] The relative change $Q_\mathrm{dyn}$ in consumer surplus and health benefits,
    \begin{align}
        Q_\mathrm{dyn}(s, G) &= \frac{1}{CC(s)} \, \sum_\omega \sum_{e \in s} \left[ \zeta(\omega) \, \left(\beta \, n(\omega,G)(P(\omega, G)-1) \, \frac{\tau(\omega, G_\mathrm{base})-\tau(\omega, G)}{2} + \frac{n(\omega, G_\mathrm{base})+n(\omega, G)}{2}\right) \right. \nonumber \\
        & \quad\quad\quad\quad\quad\quad\quad  + \; \xi(\omega) \, \beta \, n(\omega,G) \, (P(\omega, G)-1) \, \ell(\omega, G) \bigg] \, \Delta \tau_e(\omega) \\
        &\approx \frac{\Delta \mathrm{CS}(s, G) + \Delta \mathrm{HB}(s, G)}{\mathrm{CC}(s)} \,,  \nonumber
    \end{align}
    representing the relative change in net present value as a result of travel time reductions and induced demand relative to the required investment costs, under the assumption that route choices remain unchanged.
\end{enumerate}
Here, we evaluate the demand $n(\omega, G)$ at time $t=0$ without population growth and neglect future maintenance costs, since the exact time of construction of the segment is unknown during the planning stage for the backward algorithm. Including population growth would only add a multiplicative factor to the importance of all segments without changing the relative ordering.

Details on the extension of the original evaluation function for individual edges to the segment-evaluation $Q_\mathrm{pen}$ are given in Supplementary Note 1. Derivations of the evaluation criteria $Q_\mathrm{stat}$ and $Q_\mathrm{dyn}$ from the net present value are given in Supplementary Note 3.

\clearpage

\section*{Data availability}
Data will be made available on request to M.P.

\section*{Code availability}

The code for the backwards percolation algorithm is available through GitHub (\url{https://github.com/PhysicsOfMobility/BikePathNet} in the dev branch) under AGPL-3.0 license.
Code for the greedy direct optimization and the evaluations of the net present value is also available through GitHub (\url{https://github.com/madspDTU/OptimalBicycleExpansion}). Code for the batched direct optimization is available upon request.

\bibliography{bibliography}

\section*{Acknowledgements}
The authors thank Henrik Wolf for significant improvements to the code for the backwards percolation approach.
C.S. acknowledges support from the Deutsche Bundesstiftung Umwelt (DBU, German Federal Environmental Foundation). 
The project was partially funded by the Deutsche Forschungsgemeinschaft (DFG, German Research Foundation) – project number 493613373 to M.S.  Both M.P. and J.R. were funded by the Danish Road Directorate’s Cycling Fund (Vejdirektoratets Cykelpulje) of 2022  project number
CP22-022.
The authors gratefully acknowledge the computing time provided on the high-performance computers at the NHR Centers NHR@TUD. This is funded by the Federal Ministry of Education and Research and the state governments participating on the basis of the resolutions of the GWK for the national high-performance computing at universities (www.nhr-verein.de/unsere-partner).

The funders had no role in study design, data collection and analysis, decision to publish or preparation of the manuscript.

\section*{Author contribution} 

C.S. conceived the research of designing the experiments and analyzing the empirical data. C.S. and M.P. performed the simulations.  All authors planned and designed the research, contributed to interpreting the results and writing the manuscript.

\section*{Additional Information}
The authors declare no competing interests.

\newpage
\section*{Supplemental Material}
\appendix
\renewcommand{\thefigure}{S\arabic{figure}}
\setcounter{figure}{0}
\renewcommand{\thetable}{S\arabic{table}}
\setcounter{table}{0}
\renewcommand{\theequation}{S\arabic{equation}}
\setcounter{equation}{0}

\subsection*{Supplementary Note 1: Bikeability with variable demand}

In the original proof-of-concept introduction of the dynamic backwards percolation approach \cite{Steinacker2022_DemandDrivenDesign}, the quality of bike path networks $G$ was evaluated with the bikeability, 
\begin{align}
    B(G) &= \frac{T(G_\mathrm{base}) - T(G)}{T(G_\mathrm{base}) - T(G_\mathrm{full})} \,
\end{align}
based on the total effective travel time in the network
\begin{align}
    T(G) &= \sum_{\omega} T(\omega, G) = \sum_{\omega} n(\omega, G)\,\tau(\omega, G)
\end{align}
compared to the maximum possible travel time $T(G_\mathrm{base})$ in the base state and the minimum possible travel time $T(G_\mathrm{full})$ in the fully built state of the network. The bikeability describes the normalized change between the two extreme states, thus offering a transferable measure for the improvement of cycling compared to its maximum potential (i.e. without discriminating against cities where cycling is difficult by default, for example due to the geography or topography).

However, with variable demand, the total travel time is not a suitable measure. The total travel time $T(\omega) = n(\omega, G) \, \tau(\omega, G)$ for one trip $\omega$ may increase even though the travel time $\tau(\omega, G)$ for a trip decreases due to induced demand increasing the number of cyclists $n(\omega, G)$. Note that we here only consider induced demand due to shorter travel times, not demand changes due to population growth, such that we work without explicit reference to time in the following.

We define a new loss function $L$, explicitly taking into account the variable demand $n(\omega, \tau)$ as a function of the effective travel time $\tau(\omega, G)$ of each trip $\omega$. As described in the Methods section in the main manuscript, the demand curve for each trip $\omega$ is given as a logit model of the form
\begin{align}
 n(\omega, \tau) &= n(\omega) \, P(\omega, \tau) = n(\omega) \frac{e^{\beta\,\tau}}{e^{\beta\,\tau} + e^{\beta\,\tau_\mathrm{other}(\omega)}} \,,
\end{align}
where we substitute the implicit dependence on the network $G$ with the explicit dependence on the travel time $\tau$ to simplify the calculations here and in Supplementary Note 3. 
The change in loss $\Delta L$ for a single trip $\omega$ should be given by the change in travel time, $\Delta L(\omega) = \Delta T(\omega) = n(\omega, \tau) \, \Delta \tau$ for a change in the network causing a change $\Delta \tau$ of the travel time. Additional effects like induced demand introduce only higher order corrections proportional to $\Delta \tau^2$ that we neglect here. With this condition, the evaluation function defined in Supplementary Note 2 [Eq.~\eqref{eq:Q_pen}] estimates this difference relative to the length of the segments, as in the original definition. For a single trip $\omega$, we thus have a loss function of the form
\begin{align}
    L(\omega, G) &= \int_{0}^{\tau(\omega, G)}n(\omega, \tau') \, \mathrm{d}\tau' \\
    &= n(\omega) \, \int_{0}^{\tau(\omega, G)} \frac{e^{\beta\,\tau'}}{e^{\beta\,\tau'} + e^{\beta\,\tau_\mathrm{other}(\omega)}} \, \mathrm{d}\tau' \\
    &= \frac{n(\omega)}{\beta} \ln{\left(\frac{e^{\beta\,\tau(\omega, G)}+ e^{\beta\,\tau_\mathrm{other}(\omega)}}{1 + e^{\beta\,\tau_\mathrm{other}(\omega)} }\right)} \,.
\end{align}

\begin{figure}[ht]
    \centering
    \includegraphics{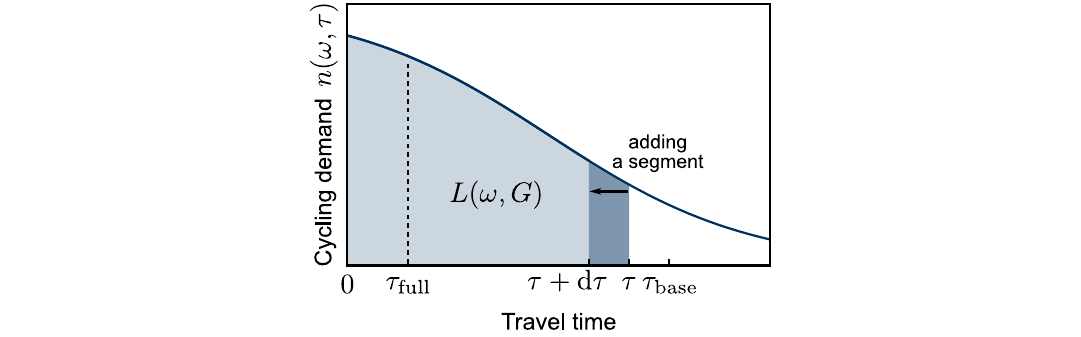}
    \caption{
        Bikeability loss function with variable demand. The loss function $L(\omega, G)$ describes the area under the demand curve (shaded area). A reduction in travel time due to an added segment (black arrow, dark blue area) reduces the loss function. In the limit of constant demand $n(\omega)$, the loss function becomes $L(\omega, G) = n(\omega)\,\tau(\omega, G) = T(\omega, G)$, equal to the total travel time of cyclists. The bikeability [Eq.~\eqref{eq:supp_bikeability}] evaluates the relative change of the loss function between the base network (large travel times $\tau_\mathrm{base}$) and the fully upgraded network (small travel times $\tau_\mathrm{full}$).
    }
    \label{fig:FIGS1_BikeabilityLossFunction}
\end{figure}

\newpage

This loss function measures the area under the demand curve $n(\tau)$ (Fig.~\ref{fig:FIGS1_BikeabilityLossFunction}). When the travel time $\tau$ becomes smaller, $\mathrm{d}\tau < 0$, the loss reduces with $\mathrm{d} L(\omega) = n(\omega, \tau) \, \mathrm{d} \tau < 0$ as required. The same travel time reduction $\mathrm{d} \tau$ has a larger effect if the travel time is already small and the number of cyclists is large. For static demand $n(\omega, \tau) = n_\mathrm{base}(\omega) = n(\omega, \tau(\omega, G_\mathrm{base}))$, this definition simplifies to the original function
\begin{align}
    L(\omega, G) &= \int_{0}^{\tau(\omega, G)} n_\mathrm{base}(\omega) \, \mathrm{d}\tau' \\
    &= n_\mathrm{base}(\omega)\,\tau(\omega, G) \\
    &= T(\omega, G) \,,
\end{align}
measuring the total travel time.

For multiple trips, the total loss function of the network is simply the sum of the loss over all trips,
\begin{align}
    L(G) &= \sum_{\omega} L(\omega, G) \,.
\end{align}

To compute the overall bikeability of the network, we employ the same idea of comparing the loss function of the network with the initial and the fully built network,
\begin{align}
    B(G) &= \frac{L\left(G_\mathrm{base}\right) - L\left(G\right)}{L\left(G_\mathrm{base}\right) - L\left(G_\mathrm{full}\right)} \,. \label{eq:supp_bikeability}
\end{align}
The bikeability $B(G) = 0$ in the initial network and increases as more bike paths are built, up to $B(G) = 1$ in the fully upgraded network [Eq.~(4) in the main manuscript].

\newpage

\subsection*{Supplementary Note 2: Consistent evaluation of bike link importance}

In the original proof-of-concept introduction of the dynamic backwards percolation approach \cite{Steinacker2022_DemandDrivenDesign}, the importance of a single edge $e$ in the bike path network was estimated by the product of the number of cyclists $n_e$ using that edge and the penalty $c_e = \frac{v_\mathrm{csh}}{v_e} > 1$ on the travel time of the edge when the bike path was removed, where $v_\mathrm{csh}$ and $v_e$ denote the speed of cyclists on the cycle superhighway edge and the base network edge, respectively. Together, the importance measure is
\begin{align}
    Q_\mathrm{pen}(e) &= n_e \, c_e = \frac{n_e \, c_e \, l_e}{l_e} \,.
\end{align}
This measure is proportional to the total travel time $T_e = n_e\,c_e\,\frac{l_e}{v_{e,0}}$ on the edge per unit length of the edge. 

However, the cycle superhighway network in the main manuscript consists of segments which each include multiple edges. We now define an evaluation function across multiple edges at the same time to extend the penalty-based evaluation. For consistency, the evaluation function should fulfill two general conditions: 
\begin{enumerate}
    \item[(i)] a segment consisting of multiple identical edges with the same number of cyclists should be assigned the same importance as the individual edges
    \item[(ii)] a segment consisting of a single edge should be assigned the same importance as the individual edge
\end{enumerate}

\begin{figure}[ht]
    \centering
    \includegraphics{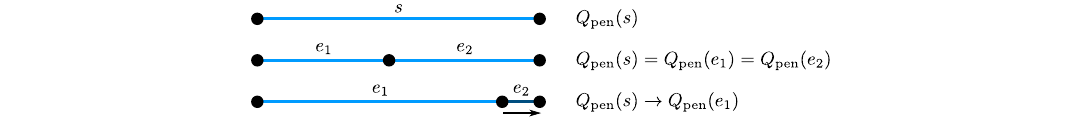}
    \caption{
        Consistent definition of segment importance. The definition of importance for edges and segments consisting of multiple edges should fulfil consistency conditions such that the resulting planned networks do not depend on the formatting of input data. In particular, the importance of an edge or a segment (top) should not change if it is split into multiple edges (middle) and should not be affected by edges with negligible length (bottom).
    }
    \label{fig:FIGS2_PenaltyModel}
\end{figure}

Condition (i) ensures that the order in which edges are removed does not depend on the fine-graining of edges in the network. For example, splitting edges and adding additional nodes or merging edges in a segment to a single edge should not change the importance. This condition also already ensured that the original definition was independent of the fine-graining of the network. Condition (ii) ensures that the definition for segments is consistent with the original evaluation function in the limiting case of a segment consisting of a single edge.

\newpage

These two conditions are fulfilled by a length-weighted sum of the contribution of the individual edges in a segment [Eq.~(17) in the main manuscript],
\begin{align}
    Q_\mathrm{pen}(s, G) &= \frac{\sum_\omega \sum_{e \in s} n_e(\omega, G) \, l_e \, c_e(\omega)}{\sum_{e \in s} l_e} \,. \label{eq:Q_pen}
\end{align}
where the travel time penalty now depends on the infrastructure type of the edge and the cyclist type of the trip.

This evaluation function again describes the total added travel time $\Delta T$ along all edges of the segment relative to the total length of the segment and thus naturally reduces to the original version when considering only a single edge. Here, we evaluate the importance of the segment with respect to the demand $n(\omega, G)$ in the current network, thus ignoring any additional induced demand due to already built bike paths.

To check the first condition, consider $n$ cyclists from a single trip $\omega$ crossing a segment $s = (e_1,e_2)$ with two identical edges, $c_1 = c_2 = c$, with arbitrary length. The measure is identical to the original definition for a single edge,
\begin{align}
    Q_\mathrm{pen}(s) &= \frac{n \, l_1 \, c_1 + n \, l_2 \, c_2}{l_1 + l_2} \\
    &= \frac{n \, c \, \left(l_1 + l_2\right)}{l_1 + l_2} \\
    &= n \, c \\
    &= Q_\mathrm{pen}(e_1) = Q_\mathrm{pen}(e_2)
\end{align}
since the length of the segment cancels out due to the identical parameters. 

To explicitly check the second condition, consider $n$ cyclists from a single trip $\omega$ crossing a segment $s = (e_1,e_2)$, such that the number of cyclists on both edges is the same $n_1 = n_2 = n$. Here, the two edges may be different, $c_1 \neq c_2$, but we assume that the second edge has negligible length, $l_2 \rightarrow 0$. Then, the measure reduces to the original definition on the first edge
\begin{align}
    Q_\mathrm{pen}(s) &= \lim_{l_2 \rightarrow 0} \frac{n \, l_1 \, c_1 + n \, l_2 \,c_2}{l_1 + l_2} \\
    &= \frac{n \,c_1\, l_1}{l_1} \\
    &= n_1 \, c_1 \\
    &= Q_\mathrm{pen}(e_1)
\end{align}
since the second edge does not contribute to the importance weighted by its length.

\newpage

\subsection*{Supplementary Note 3: Derivation of the evaluation criteria for the dynamic backward percolation approach}
To achieve a fairer comparison to the direct optimization approach, we derive two additional evaluation functions for the backward percolation approach explicitly based on the net present value
\begin{align}
    \mathrm{NPV}(t) &= \mathrm{CS}(t) + \mathrm{HB}(t) - \mathrm{CC}(t) - \mathrm{MC}(t) + \mathrm{SV}(t) \,.
\end{align}
As discussed in the main manuscript, the positive effects of the consumer surplus $\mathrm{CS}$ and the health benefits $\mathrm{HB}$ dominate the net present value. Moreover, construction and maintenance costs and scrap value depend explicitly on the timing of construction, not just the ordering of construction, and are thus difficult to estimate in the backward percolation approach. Therefore, we focus on estimating the effects of individual cycle superhighway segments on the consumer surplus and the health benefits.

The consumer surplus describes the value of changes in the travel time $\tau$ due to new bike paths, computed with the rule-of-half to account for the variable demand. The health benefits describe the value of the distance $\ell$ travelled by all cyclists (compare Methods section in the main manuscript). We thus start from the definitions
\begin{align}
    \mathrm{CS}(t) &= \sum_\omega \frac{1}{2}\zeta(\omega) \left[n_\mathrm{base}(\omega) + n(\omega, \tau)\right] \left[\tau_\mathrm{base}(\omega) - \tau(\omega)\right] \\
    \mathrm{HB}(t) &= \sum_\omega \xi(\omega) \left[n(\omega, \tau) \ell(\omega) -  n_\mathrm{base}(\omega) \ell_0(\omega) \right] \,.
\end{align}
where $n(\omega, \tau)$ and $n_\mathrm{base}(\omega) = n(\omega, \tau_\mathrm{base})$ denote the number of cyclists for a trip-cyclist type-combination $\omega$ and $\tau(\omega)$ and $\tau_\mathrm{base}(\omega)$ denote their travel time in the current network $G$ and the initial network $G_\mathrm{base}$, respectively. 

In general, removing any segment $s$ of the cycle superhighway network affects the consumer surplus and the health benefits by changing the travel time $\tau$, the travel distance $\ell$, and the number of cyclists $n$. Expanding the changes $\Delta \mathrm{CS}$ and $\Delta \mathrm{HB}$ in these variables yields
\begin{align}
    \Delta \mathrm{CS} &= \mathrm{CS}(\mathrm{current\;state}) - \mathrm{CS}(\mathrm{edge\;removed}) \\
    &= \sum_\omega \frac{1}{2}\zeta(\omega) \left[\left(n_\mathrm{base}(\omega) + n(\omega, \tau)\right) \left(\tau_\mathrm{base}(\omega) - \tau(\omega)\right) - \left(n_\mathrm{base}(\omega) + n(\omega, \tau'(\omega))\right) \left(\tau_\mathrm{base}(\omega) - \tau'(\omega)\right)\right] \\
    &= \sum_\omega \frac{1}{2}\zeta(\omega) \left[- \left(\tau_\mathrm{base}(\omega) - \tau(\omega) \right) \Delta n(\omega, \tau) + \left(n_\mathrm{base}(\omega) + n(\omega, \tau) \right) \Delta \tau(\omega) + \Delta  n(\omega, \tau) \Delta\tau(\omega)\right],
\end{align}
\begin{align}
    \Delta \mathrm{HB} &= \mathrm{HB}(\mathrm{current\;state}) - \mathrm{HB}(\mathrm{edge\;removed}) \\
    &= \sum_\omega \xi(\omega) \left[ \left(n(\omega, \tau) \ell(\omega) -  n_\mathrm{base}(\omega) \ell_\mathrm{base}(\omega) \right) - \left(n(\omega, \tau'(\omega)) \ell'(\omega) -  n_\mathrm{base}(\omega) \ell_\mathrm{base}(\omega) \right)\right] \\
    &= \sum_\omega \xi(\omega) \left[ n(\omega, \tau) \ell(\omega) -  \left(n(\omega,t)+\Delta n(\omega, \tau) \right)\left( \ell(\omega)+\Delta \ell(\omega)\right)\right] \\
    &= \sum_\omega \xi(\omega) \left[- n(\omega,\tau)\Delta\ell(\omega) - \Delta n(\omega, \tau) \ell(\omega) - \Delta n(\omega,t)\Delta\ell(\omega)\right]
\end{align}
where the dashed variables denote the values for the network $G \setminus s$ in which the segment is removed, $n(\omega, \tau') = n(\omega, \tau) + \Delta n(\omega, \tau)$, $\tau'(\omega) = \tau(\omega) + \Delta \tau(\omega)$, and $\ell'(\omega) = \ell(\omega) + \Delta\ell$. 

We assume that cyclists do not change their route as a result of the change in the network,
\begin{align}
    \Delta\ell &= 0 \, .
\end{align}
This simplification is the same as for the original evaluation function \cite{Steinacker2022_DemandDrivenDesign} and strongly reduces computation times by avoiding recalculation of routes to evaluate the importance of each segment in the network in every step. The change in travel time $\Delta \tau(\omega)$ is then given by the sum of all changes in travel time along each edge included in the highway segment $s$
\begin{align}
    \Delta \tau(\omega) = \sum_{e \in s} \Delta \tau_e(\omega) = \sum_{e \in s} \frac{l_e}{v_e(\omega)} - \frac{l_e}{v_\mathrm{csh}(\omega)} = \frac{l_e}{v_\mathrm{csh}(\omega)} \, \Delta c_e(\omega) \, ,
\end{align}
where $l_e$ is the length of edge $e$ and $\Delta c_e(\omega) = c_e(\omega) - 1 = \frac{v_\mathrm{csh}(\omega)}{v_e(\omega)} - 1$ is the change of the travel time penalty on the edge. Here, again, $v_\mathrm{csh}(\omega)$ denotes the speed of cyclists on a cycle superhighway and $v_e(\omega)$ denotes the speed of cyclists on the base infrastructure type of the edge. Finally, the number $n(\omega, \tau)$ of cyclists is not an independent function but depends on the travel time $\tau(\omega)$. We thus express the change in demand as a function of the travel time reduction, linearizing the expression
\begin{align}
\Delta n(\omega, \tau) &= n(\omega, \tau') - n(\omega, \tau) \\
&= n(\omega) \, P(\omega, \tau') - n(\omega) \,P(\omega, \tau) \\
&= n(\omega) \, \left(\frac{e^{\beta\,\tau'}}{e^{\beta\,\tau'} + e^{\beta\,\tau_\mathrm{other}}} -\frac{e^{\beta\,\tau}}{e^{\beta\,\tau} + e^{\beta\,\tau_\mathrm{other}}} \right) \\
&= n(\omega) \, \left(\frac{e^{\beta\,\tau+\Delta\tau}}{e^{\beta\,\tau+\Delta\tau} + e^{\beta\,\tau_\mathrm{other}}} - \frac{e^{\beta\,\tau}}{e^{\beta\,\tau} + e^{\beta\,\tau_\mathrm{other}}} \right) \\
&\approx n(\omega) \, \left(\frac{e^{\beta\,\tau}}{e^{\beta\,\tau} + e^{\beta\,\tau_\mathrm{other}}} + \left(\frac{\beta e^{\beta\,\tau}}{e^{\beta\,\tau} + e^{\beta\,\tau_\mathrm{other}}} - \frac{\beta \left(e^{\beta\,\tau}\right)^2}{\left(e^{\beta\,\tau} + e^{\beta\,\tau_\mathrm{other}}\right)^2}\right) \Delta\tau(\omega) - \frac{e^{\beta\,\tau}}{e^{\beta\,\tau} + e^{\beta\,\tau_\mathrm{other}}}\right) \\
&= n(\omega) \,  \left(\beta \frac{e^{\beta\,\tau}}{e^{\beta\,\tau} + e^{\beta\,\tau_\mathrm{other}}} - \beta \left(\frac{e^{\beta\,\tau}}{e^{\beta\,\tau} + e^{\beta\,\tau_\mathrm{other}}}\right)^2\right) \Delta\tau(\omega) \\
&= \beta\,n(\omega) \left( P(\omega, \tau) - P^2(\omega, \tau) \right) \Delta\tau(\omega) \\
&= \beta n(\omega, \tau) \left(1-P(\omega, \tau)\right)\Delta\tau(\omega) \\
&= \sum_{e \in s} \beta n(\omega, \tau) \left(1-P(\omega, \tau)\right) \, \Delta \tau_e(\omega) \,,
\end{align}
where the last line gives the expression for a segment $s$ as the sum over all edges $e$ in the segment.

We now define the evaluation function $Q_\mathrm{dyn}$ as the linearized change in consumer surplus and health benefits relative to the construction costs of the segment,
\begin{align}
    Q_\mathrm{dyn}(s) &\approx \frac{\Delta \mathrm{CS}(s) + \Delta \mathrm{HB}(s)}{CC(s)} \\
     &\approx \frac{\sum_\omega \frac{1}{2}\zeta(\omega) \left[- \left(\tau_\mathrm{base}(\omega) - \tau(\omega) \right) \Delta n(\omega, \tau) + \left(n_\mathrm{base}(\omega) + n(\omega, \tau) \right) \Delta \tau(\omega)\right] + \xi(\omega) \left[-\Delta n(\omega, \tau) \ell(\omega) \right]}{CC(s)} \\
    Q_\mathrm{dyn}(s) &= \frac{1}{CC(s)} \, \sum_\omega \sum_{e \in s} \bigg[ \zeta(\omega) \left(\beta n(\omega,t)\left(P(\omega, \tau)-1\right)\frac{\tau_\mathrm{base}-\tau(\omega)}{2} + \frac{n_\mathrm{base}+n(\omega, \tau)}{2}\right) \nonumber\\
    & \quad\quad\quad\quad\quad\quad\quad\quad + \xi(\omega) \beta n(\omega,t) (P(\omega, \tau)-1) \ell(\omega)\bigg] \, \Delta \tau_e(\omega)\,. \label{eq:Q_dyn}
\end{align}

As a further simplification, we assume locally static demand, setting $\Delta n(\omega,t) = 0$ but still evaluating all expression with the current, variable demand. Only the middle term in the evaluation function Eq.~\eqref{eq:Q_dyn} remains, giving
\begin{align}
    Q_\mathrm{stat} &\approx \frac{\Delta \mathrm{CS}(s)}{CC(s)} \\
    &\approx \frac{\sum_\omega \zeta(\omega) \frac{n_\mathrm{base}(\omega) + n(\omega, \tau)}{2} \Delta \tau(\omega)}{CC(s)} \\
    Q_\mathrm{stat} &= \frac{\sum_\omega \sum_{e \in s} \zeta(\omega) \frac{n_\mathrm{base}(\omega) + n(\omega, \tau)}{2} \,\Delta \tau_e(\omega)}{CC(s)} \,. \label{eq:Q_stat}
\end{align}

\newpage

\subsection*{Supplementary Note 4: Performance overview}
\begin{figure}[ht]
    \centering
    \includegraphics[width=\textwidth]{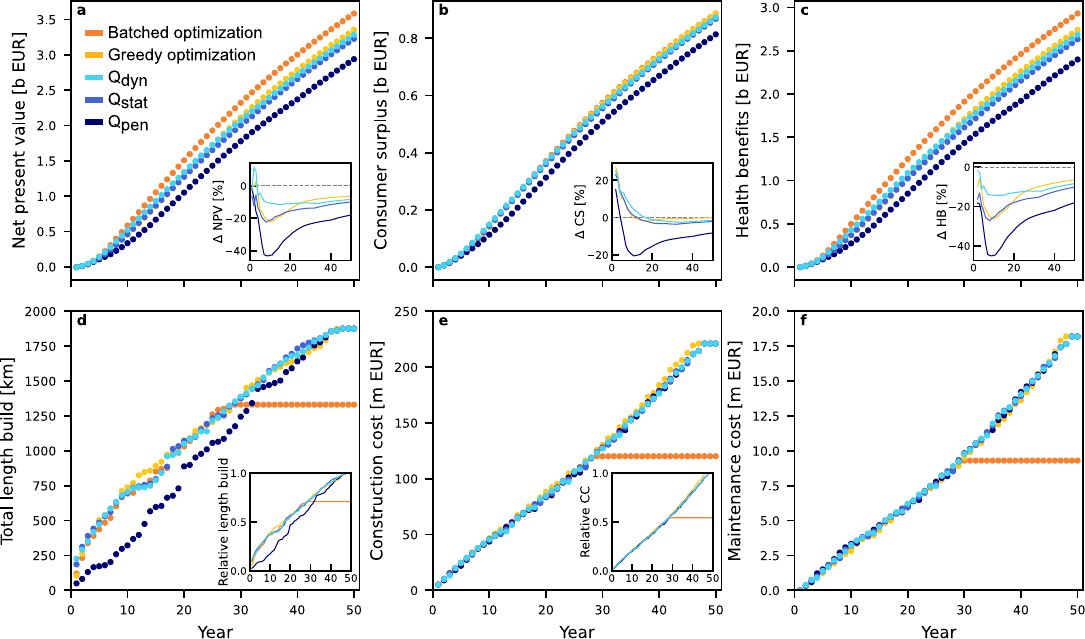}
    \caption{Performance overview for all five approaches. (a) Net present value (b) consumer surplus (CS) and (c) health benefits (HB) with (insets a-c) the relative difference to the batched optimization. (d) Total length build with relative length build in the inset, (e) total construction costs (CC) with relative construction costs in the inset and (f) maintenance costs per year. Net present value, consumer surplus and health benefits are calculated as defined in the main manuscript (compare Eq.~13, Eq.~8 and Eq.~9 respectively). The construction and maintenance costs are only the simple sum of the costs of the built segments so far, discarding any discount factors.
    }
    \label{fig:FIGS3_Performance_Overview}
\end{figure}

\newpage

\subsection*{Supplementary Note 5: Robustness of planned network expansion}
\begin{figure}[ht]
    \centering
    \includegraphics[width=0.8\textwidth]{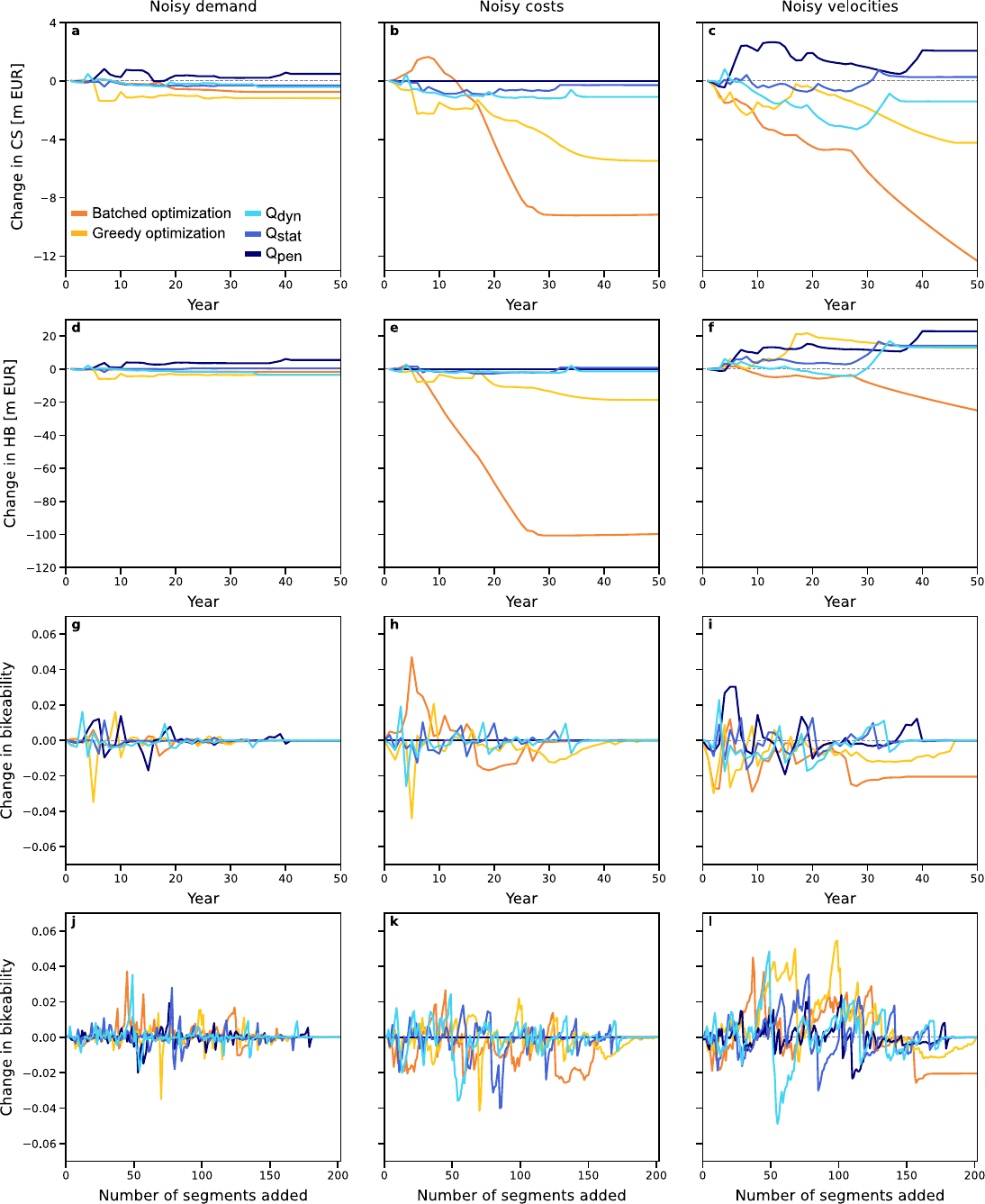}
    \caption{
        Changes of network quality measures under noisy input. Results for network quality measures [consumer surplus (top), bikeability by year (center), bikeability by segment (bottom)] to noise demand input (left), noisy segment costs (middle), and noisy velocities (right) are qualitatively similar to the results for the net present value (compare Fig.~7 in the main manuscript). Each line shows the average change over $10$ realization compared to exact input data.\vspace{-1cm}
    }
    \label{fig:FIGS4_Robustness_Metrics_Mean}
\end{figure}

\begin{figure}[ht]
    \centering
    \includegraphics[width=0.8\textwidth]{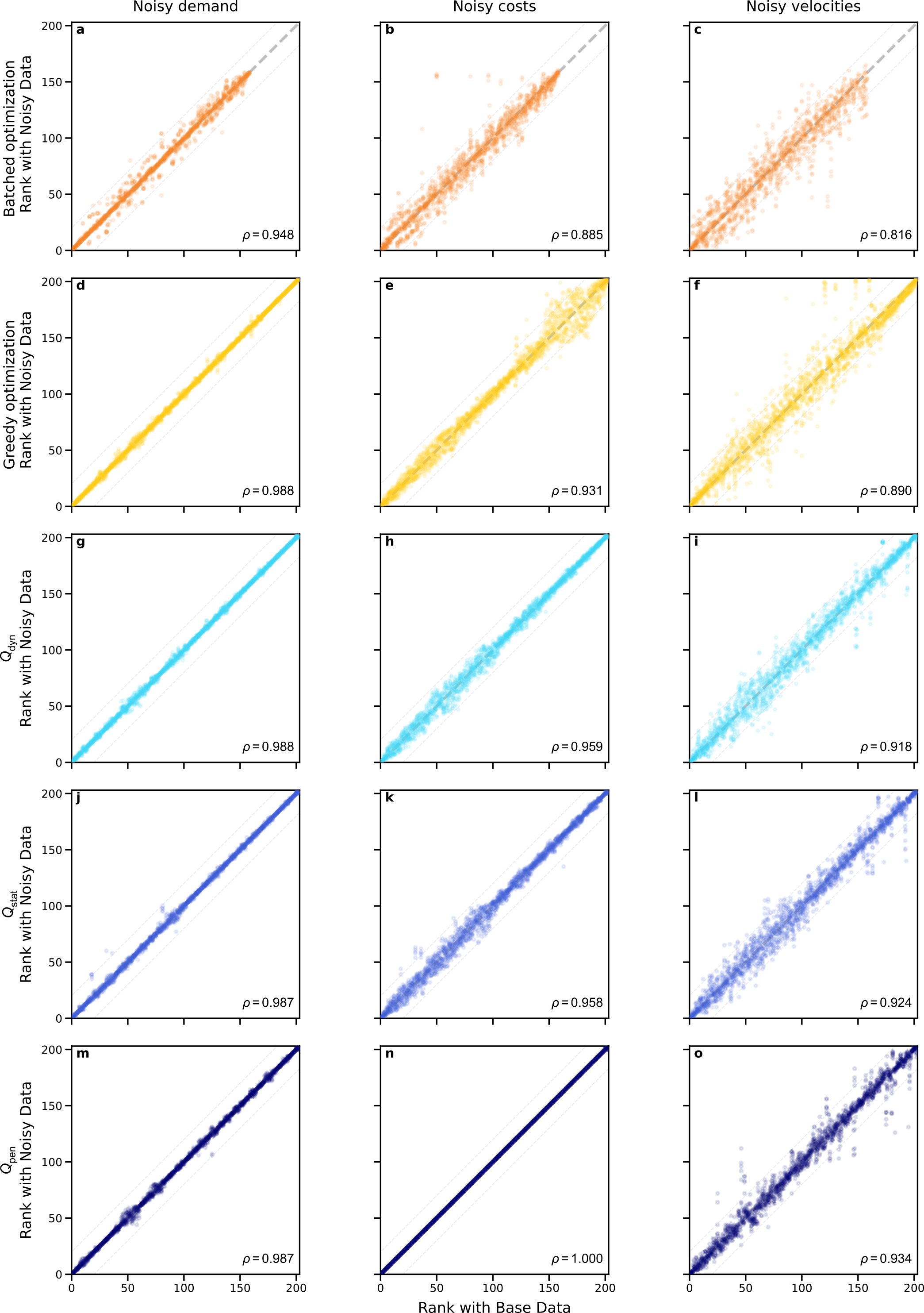}
    \caption{
        Rank differences of segment construction under noisy input data. Data on the identity indicates no change in rank. Kendall rank correlation coefficients $\rho$ (bottom right in each panel) confirm the qualitative differences observed in the main manuscript. Complex evaluation functions [(greedy) direction optimization (top) or $Q_\mathrm{dyn}$ (middle)] consistently have a larger impact on the segment ordering, whereas simpler evaluation functions [$Q_\mathrm{stat}$ and $Q_\mathrm{pen}$ (bottom)] tend to be more robust. Noisy global parameters like velocities have a larger impact than noisy distributed parameters like demand, where effects may average out over the network. The panels show rank data for $10$ realizations of noisy input each.
    }
    \label{fig:FIGS5_Robustness_Ranks}
\end{figure}

\begin{figure}[ht]
    \centering
    \includegraphics[width=\textwidth]{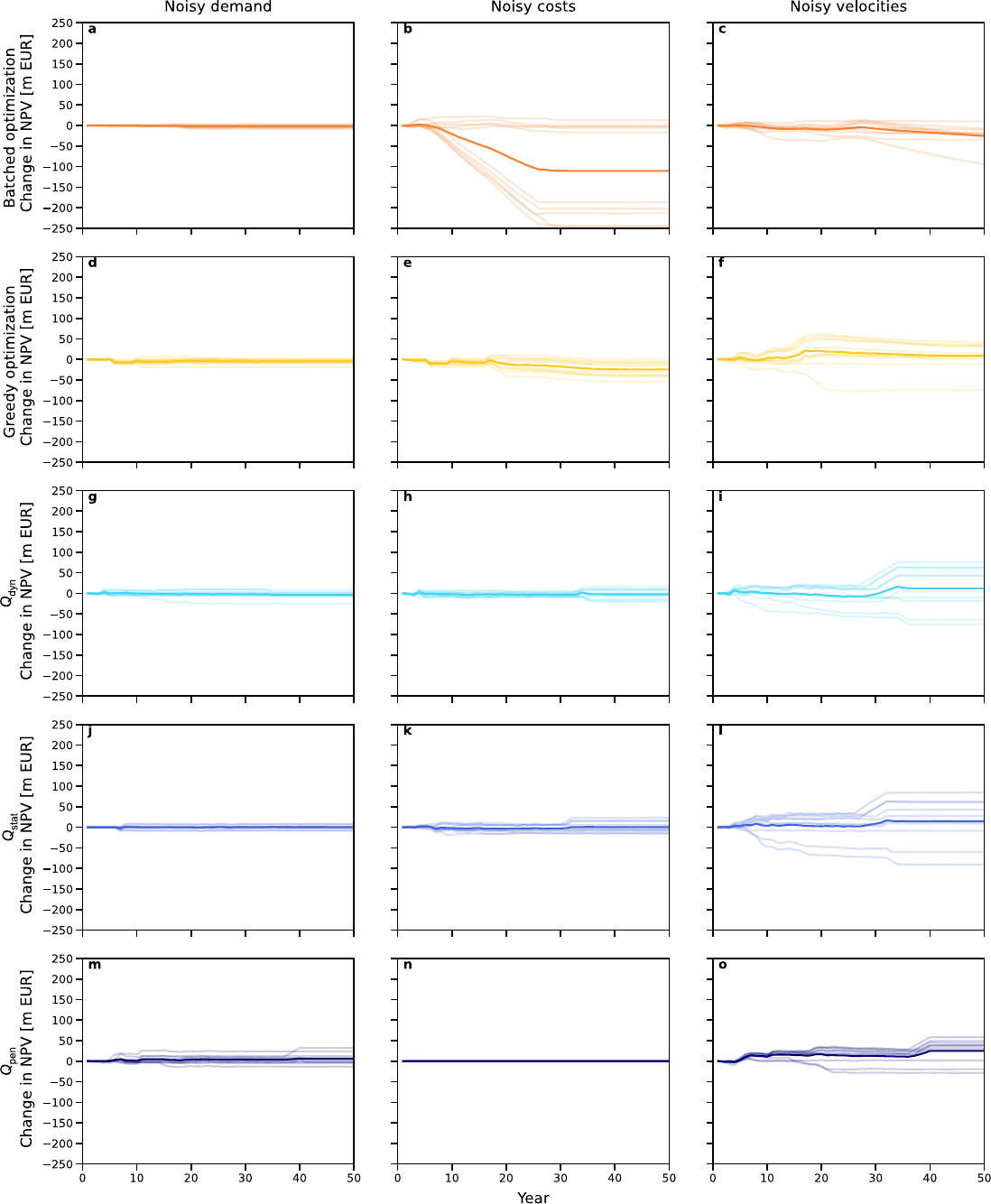}
    \caption{
        Change in net present value under noisy input data. Each panel shows $10$ independent realization (light lines) and the average (dark line).
    }
    \label{fig:FIGS6_Robustness_NPV_all}
\end{figure}

\begin{figure}[ht]
    \centering
    \includegraphics[width=\textwidth]{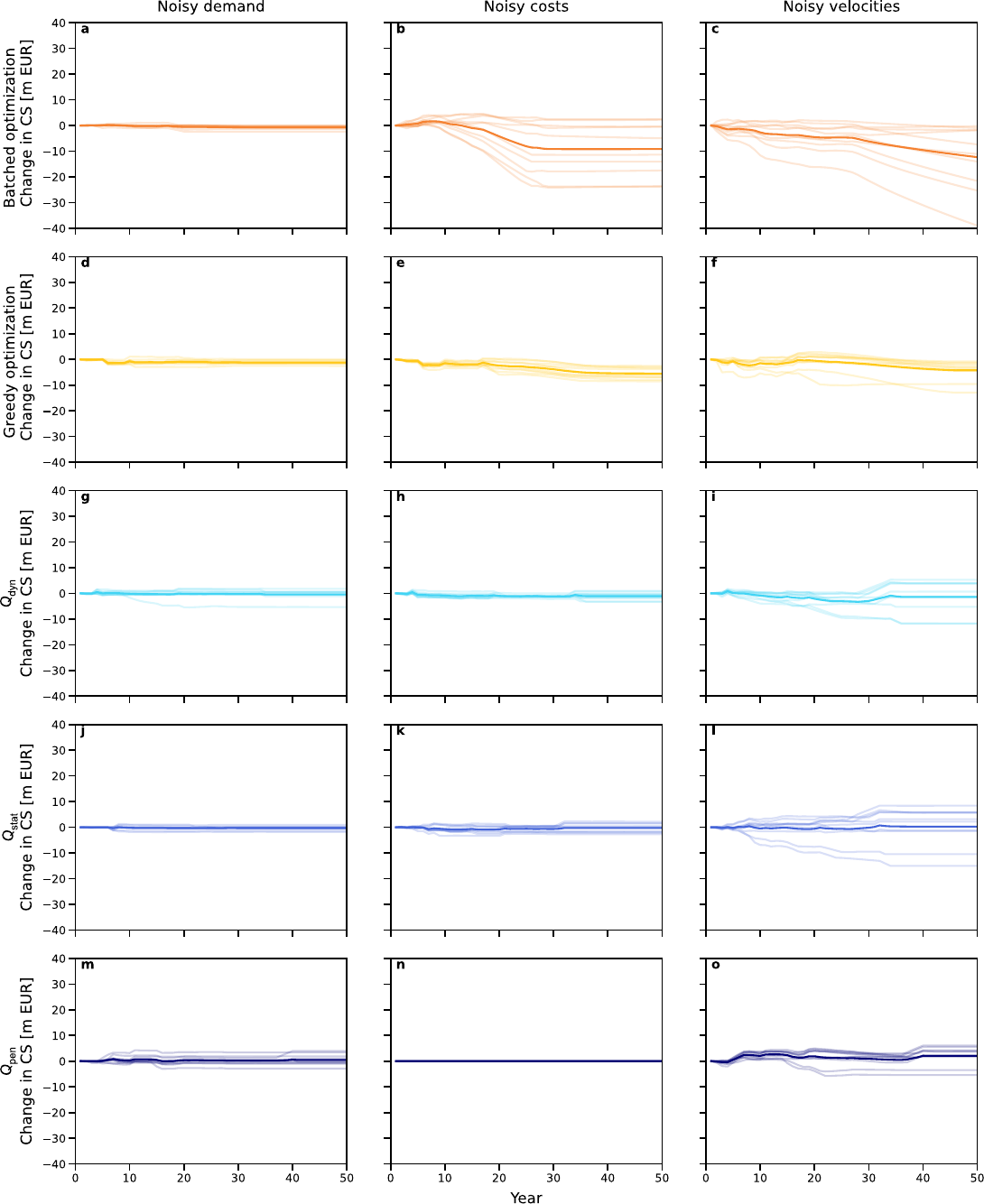}
    \caption{
        Change in consumer surplus under noisy input data. Each panel shows $10$ independent realization (light lines) and the average (dark line).
    }
    \label{fig:FIGS7_Robustness_CS_all}
\end{figure}

\begin{figure}[ht]
    \centering
    \includegraphics[width=\textwidth]{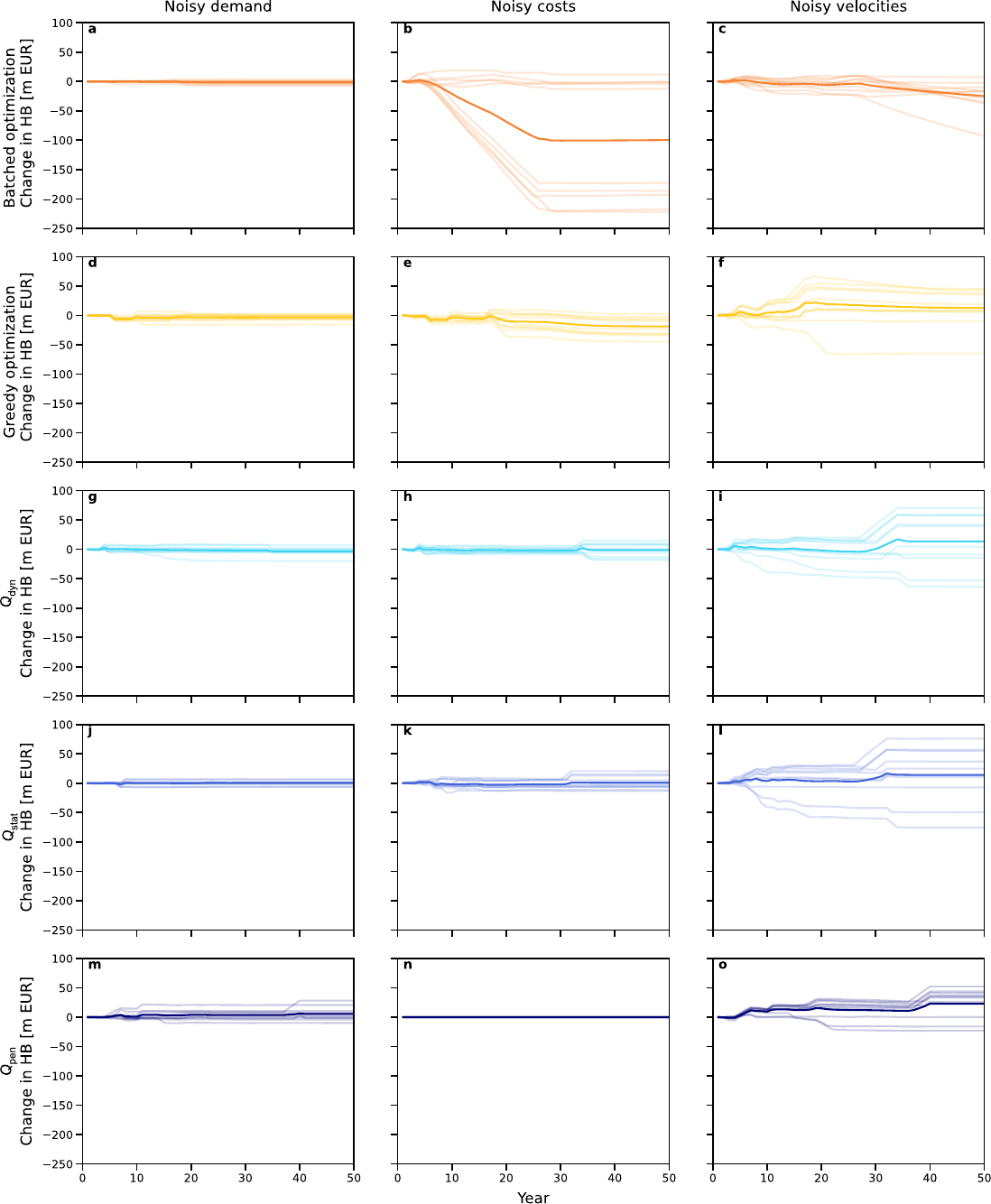}
    \caption{
        Change in health benefits under noisy input data. Each panel shows $10$ independent realization (light lines) and the average (dark line).
    }
    \label{fig:FIGS8_Robustness_HB_all}
\end{figure}

\begin{figure}[ht]
    \centering
    \includegraphics[width=\textwidth]{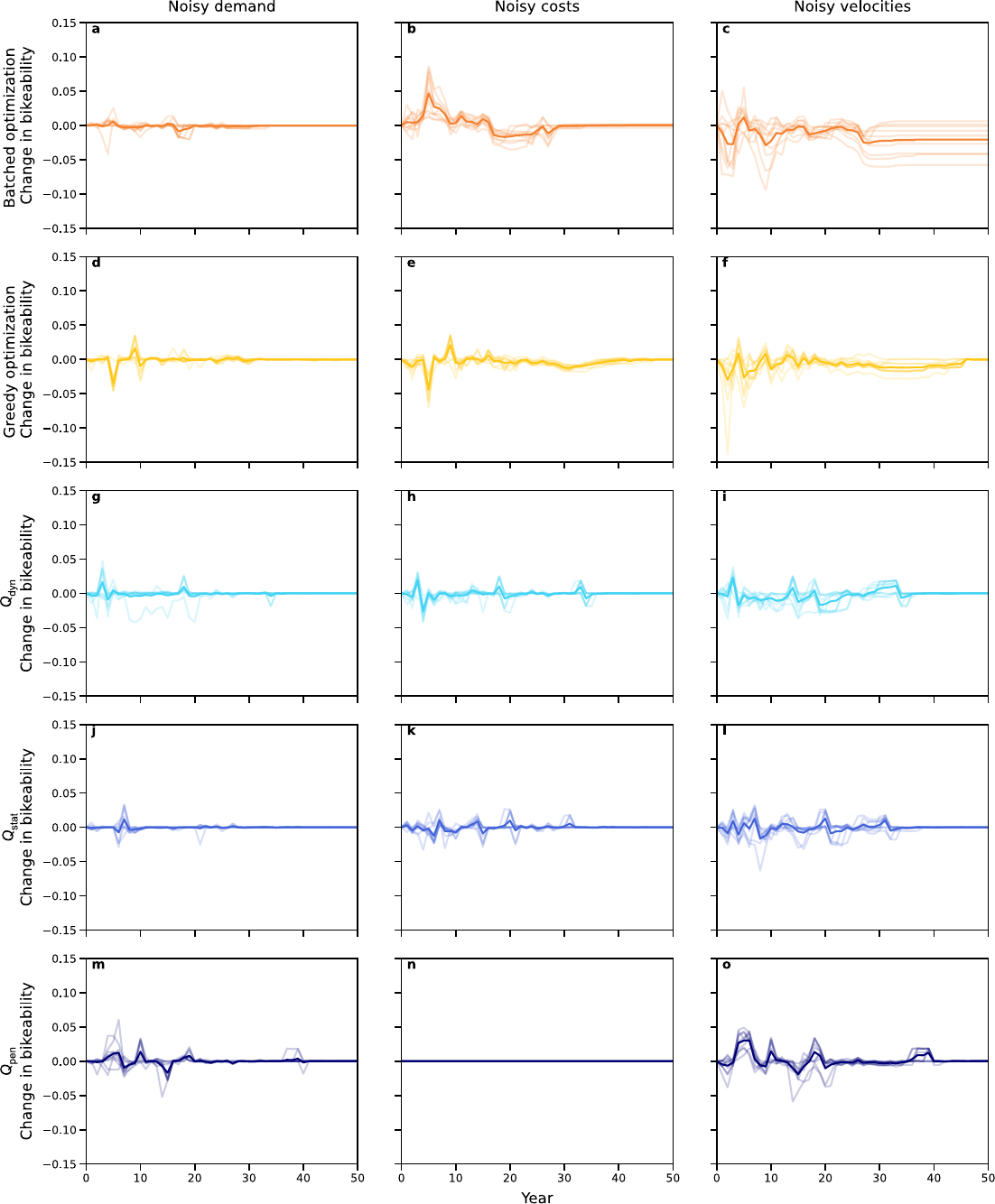}
    \caption{Change in bikeability per year under noisy input data. Each panel shows $10$ independent realization (light lines) and the average (dark line).
    }
    \label{fig:FIGS9_Robustness_BaY_all}
\end{figure}

\begin{figure}[ht]
    \centering
    \includegraphics[width=\textwidth]{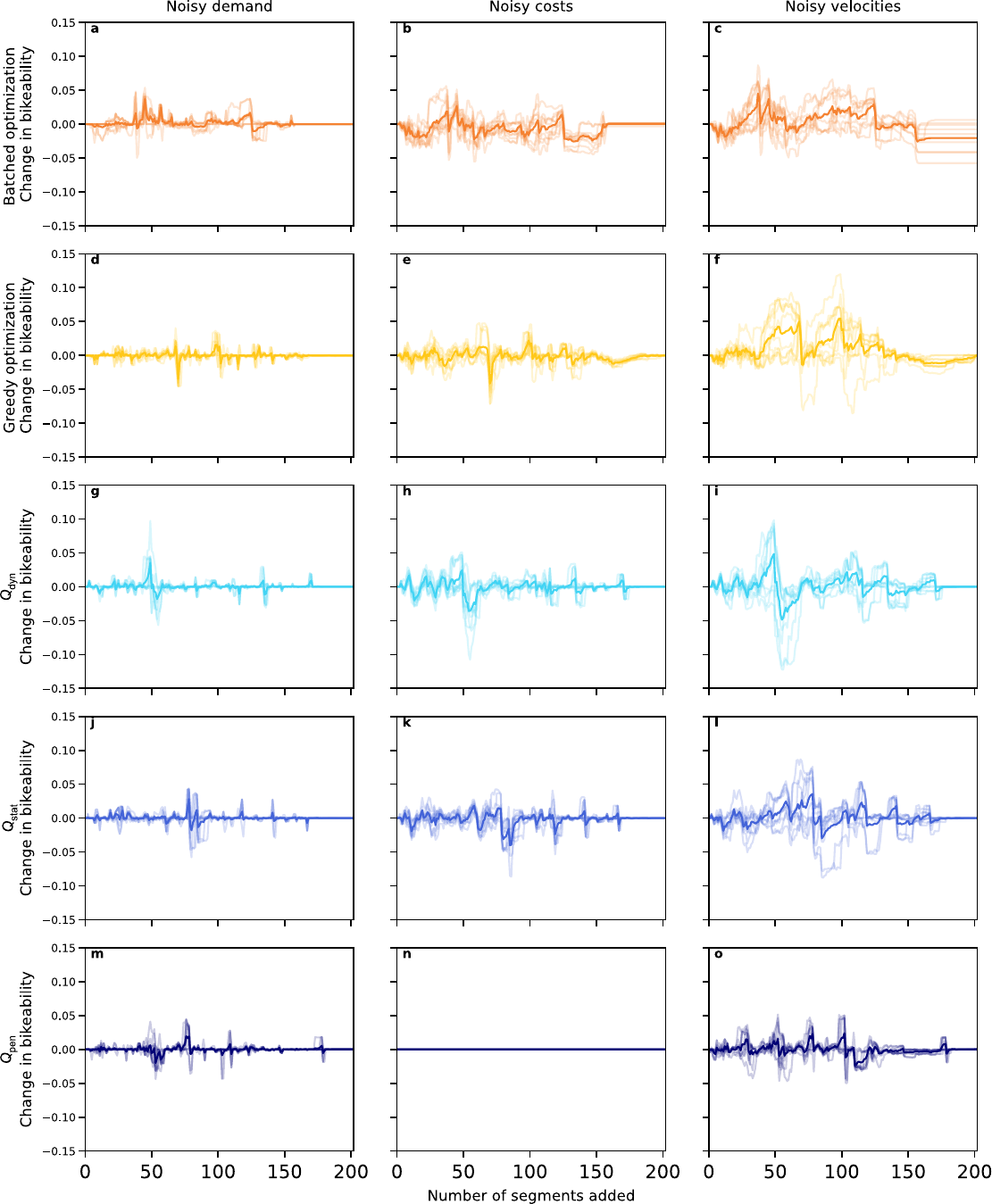}
    \caption{
        Change in bikeability per segment under noisy input data. Each panel shows $10$ independent realization (light lines) and the average (dark line).
    }
    \label{fig:FIGS10_Robustness_BaS_all}
\end{figure}

\end{document}